\documentclass[APS,twocolumn]{USG} 

\usepackage{anyfontsize} %

\usepackage{colortbl}   
\usepackage{xcolor}     

\usepackage{steinmetz}

\graphicspath{{./images/}}

\usepackage{graphicx}
\usepackage{subfig}
\usepackage{parskip}
\usepackage{color}
\usepackage{soul}
\usepackage{amssymb}

\articletype{RESEARCH ARTICLE}%

\received{XX Month Year}
\revised{XX Month Year}
\accepted{XX Month Year}
\journal{Advanced Functional Materials}
\volume{0}
\copyyear{Year}
\startpage{1}
\articledoi{10.1002/0000}

\begin{document}

\title{Artificial Spin-Glass of Antiferromagnetic Co$_3$O$_4$ Nanocubes Using Spin-Polarized Oxygen}


\author[1]{Gurleen K. Uppal}
\author[1]{Alberto Acosta}
\author[2]{Rachel Nickel}[https://orcid.org/0000-0002-8179-6295]
\author[3]{Christoph Klewe}[https://orcid.org/0000-0002-5816-5647]
\author[4]{Yanna Chen}[https://orcid.org/0000-0001-7937-4395]
\author[5]{Chengjun Sun}[https://orcid.org/0000-0001-9158-8337]
\author[6]{Charles A. Roberts}[https://orcid.org/0000-0003-4268-8379]
\author[1,7]{Johan van Lierop}[https://orcid.org/0000-0001-9398-4835]

\authormark{UPPAL \textsc{et al.}}
\titlemark{Artificial spin-glass of antiferromagnetic Co$_3$O$_4$ nanocubes using spin-polarized oxygen}

\address[1]{\orgdiv{Department of Physics and Astronomy, }\orgname{University of Manitoba, }%
\orgaddress{\state{Winnipeg, Manitoba, R3T 2N2}\country{Canada}}}

\address[2]{\orgdiv{Electronic Structure, Magnetism and Dynamics Group, }\orgname{European Synchrotron Radiation Facility, }%
\orgaddress{\state{Grenoble, 38000 Grenoble, }\country{France}}}

\address[3]{\orgdiv{Advanced Light Source, }\orgname{Lawrence Berkeley National Laboratory, }%
\orgaddress{\state{Berkeley, California 94720, }\country{United States}}}

\address[4]{\orgdiv{CLS@APS, }\orgname{Canadian Light Source, }%
\orgaddress{\state{Saskatoon, Saskatchewan, S7N 2V3, }\country{Canada}}}

\address[5]{\orgdiv{Advanced Photon Source, }\orgname{Argonne National Laboratory, }%
\orgaddress{\state{Argonne, Illinois 60439, }\country{United States}}}

\address[6]{\orgdiv{Research and Development, }\orgname{Toyota Motor Engineering and Manufacturing North America Inc., }%
\orgaddress{\state{Ann Arbor, Michigan 48105, }\country{United States}}}

\address[7]{\orgdiv{Manitoba Institute for Materials, }\orgname{University of Manitoba, }%
\orgaddress{\state{Winnipeg, Manitoba, R3T 2N2}\country{Canada}}}

\corres{Gurleen K. Uppal  (\email{uppalg1@myumanitoba.ca}) ~|~ Johan van Lierop (\email{Johan.van.Lierop@umanitoba.ca})}






\abstract[ABSTRACT]{
Preferentially exposing (100) surface terminations of Co$_{3}$O$_{4}$ using cube-shaped nanoparticles yields an unconventional rearrangement of ions that generates spin-glass-like magnetism from the antiferromagnetic nanocubes. The necessary magnetic frustration arises out of competition between polarized Co-O bonds from charge transfer between O 2p and Co 3d orbitals, and highly anisotropic Co$^{2+}$ ions combined with anomalous high spin Co$^{3+}$ ions in octahedral sites. Size control permits tuning this novel combination of ions via preferentially polarizing Co-O bonds.  The presence of control over frustrated magnetism among antiferromagnetic nanocubes yields an artificial spin glass platform for exploring unconventional nanomagnetism and engendering new opportunities for emerging technologies like spintronics and physical reservoir computing.
}




\maketitle


\section{Introduction\label{sec:intro}}

Nanoscale antiferromagnets present behaviour distinct from their bulk counterparts, including superparamagnetism-like spin dynamics that is understood to come from uncompensated surface spins that arise from broken surface coordination and disrupted magnetic exchange.  Understanding the origin of these finite-size magnetization dynamics from surface magnetism is essential for applications that exploit the absence of stray fields that antiferromagnets offer.  For example, antiferromagnetic nanoparticles used in biomedical applications like targeted drug delivery\cite{Wesselinowa.2010, Yathindranath.2011}, MRI contrast agents\cite{VanRoosbroeck.2014, Dennis.2015} and magnetic hyperthermia\cite{Demirci.2019} are much less likely to remain magnetized by the necessary applied magnetic field and clump, reducing the likelihood of unwanted uptake by the organs \cite{Vangijzegem2023SPION} or immune response.  Nanoscale antiferromagnets are increasingly sought after for their spintronics application potential.  Being insensitive to stray fields provides the option for dense packing with a significantly reduced impact from external magnetic fields.  Perhaps most exciting is the potential for ultrafast dynamics in the terahertz regime in antiferromagnet-based spintronic devices\cite{Liu.2025}.

Since the majority of antiferromagnets are based on oxides\cite{Coey.2019}, we turn to the canonical spinel antiferromagnet Co$_3$O$_4$ to explore how quantum confinement effects perturb the surface magnetism, spin dynamics and exchange interactions.  To definitively identify the physics underlying these processes, we focus on nanocubes of Co$_3$O$_4$. The surface of cubes exposes preferentially the (100) plane\cite{Su.2014, Shepit.2020}, allowing a more exact link between Co ion coordination (surface and core environments), (super) exchange interaction pathways and the  measured nanomagnetism.  Using 10, 20 and 400~nm sized cubes, we find that surface-modified Co 3d orbitals form from oxygen vacancy and core-hole formations in the O 2p band.  These surface-modified Co ion environments create frustrated spins resulting in spin-glass-like nanomagnetism.  We present a way to stabilize and control the surface spins of an antiferromagnet by way of polarized charge transfer that create metal ion (here Co$^{3+}$)-to-O 2p hybridized bonds with highly (orbital) anisotropic free spin ions (here Co$^{2+}$ and high-spin Co$^{3+}$).

\section{Results and Discussion\label{sec:results}}

\subsection{Cube composition and morphology}


\begin{figure}[t!]

\centering

\subfloat[a)\label{fig:XRD}]{\includegraphics[width=0.4\textwidth]{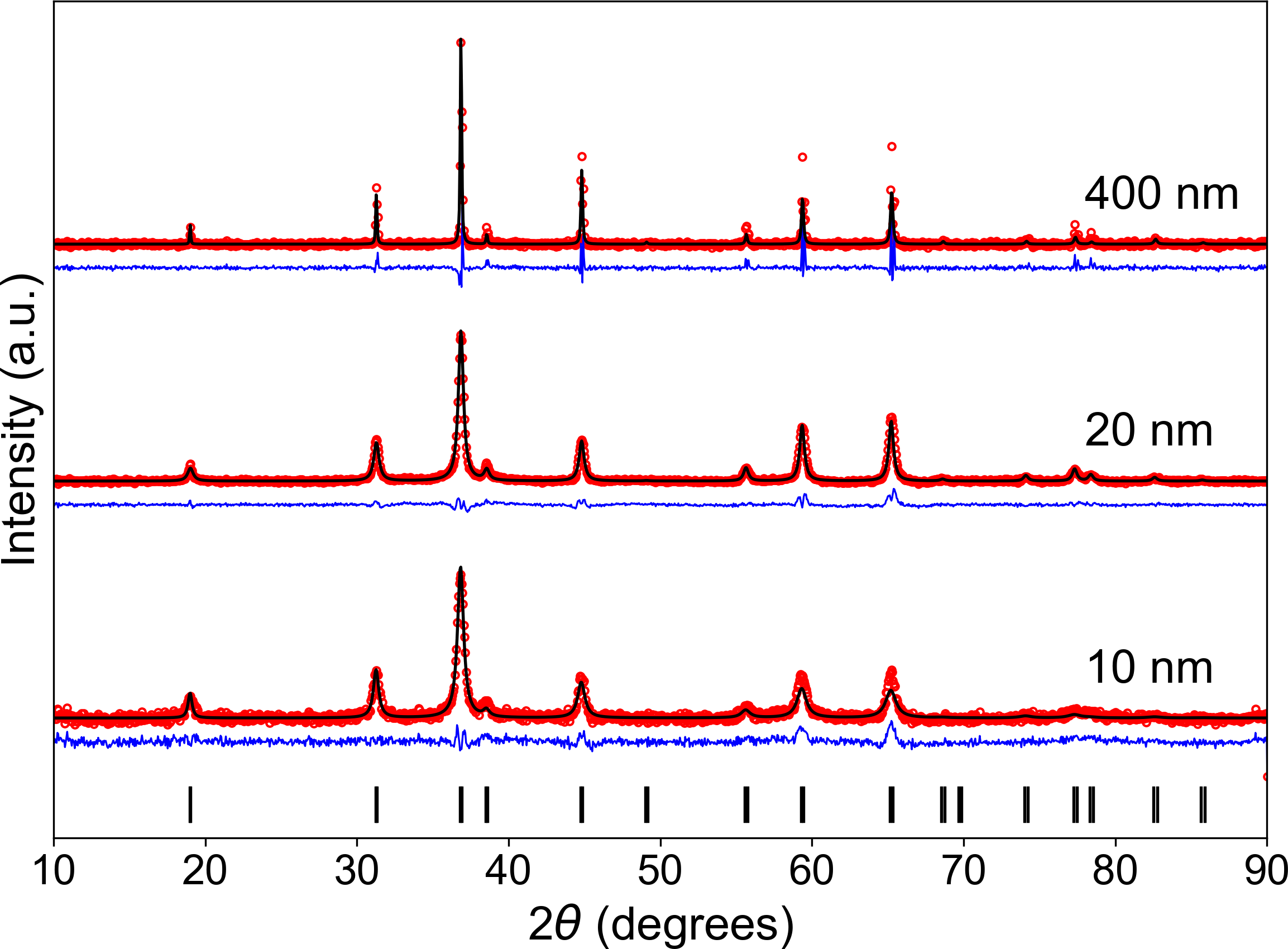}}

\subfloat[b) \label{fig:TEM_10nm}]{\includegraphics[width=0.2\textwidth]{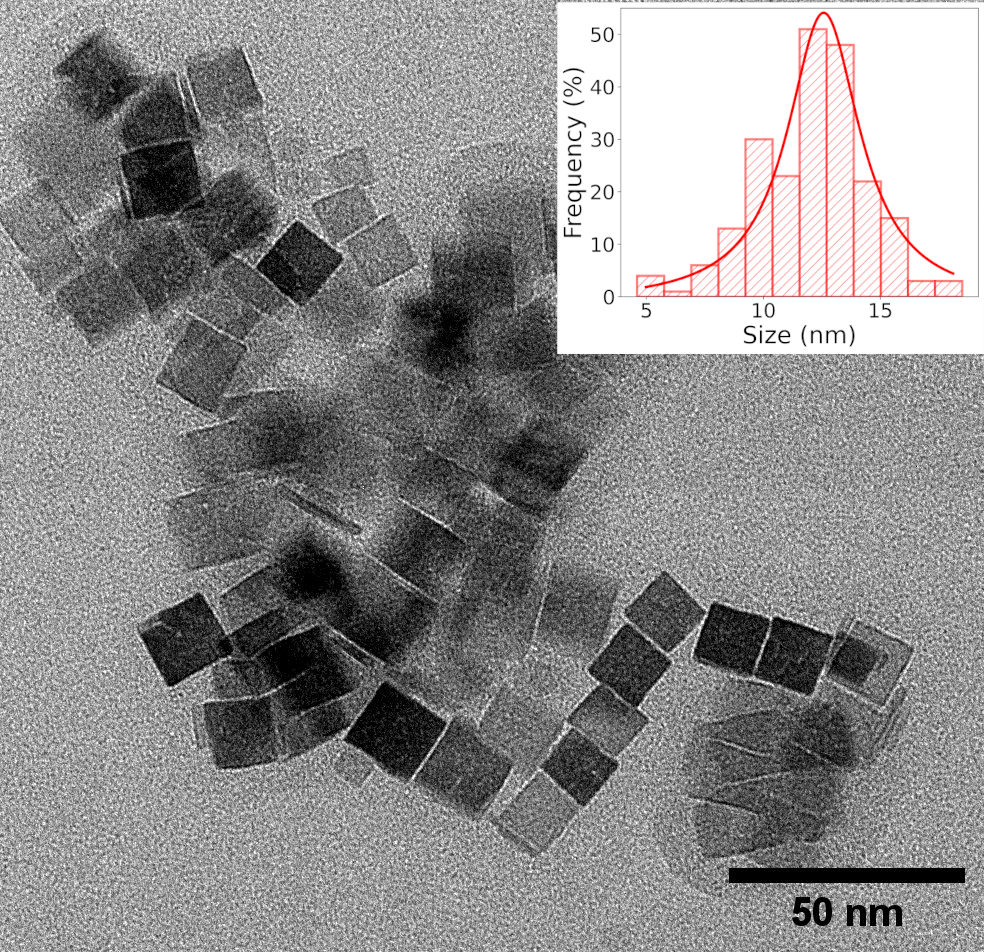}}
\hspace*{6pt}
\subfloat[c) \label{fig:TEM_20nm}]{\includegraphics[width=0.2\textwidth]{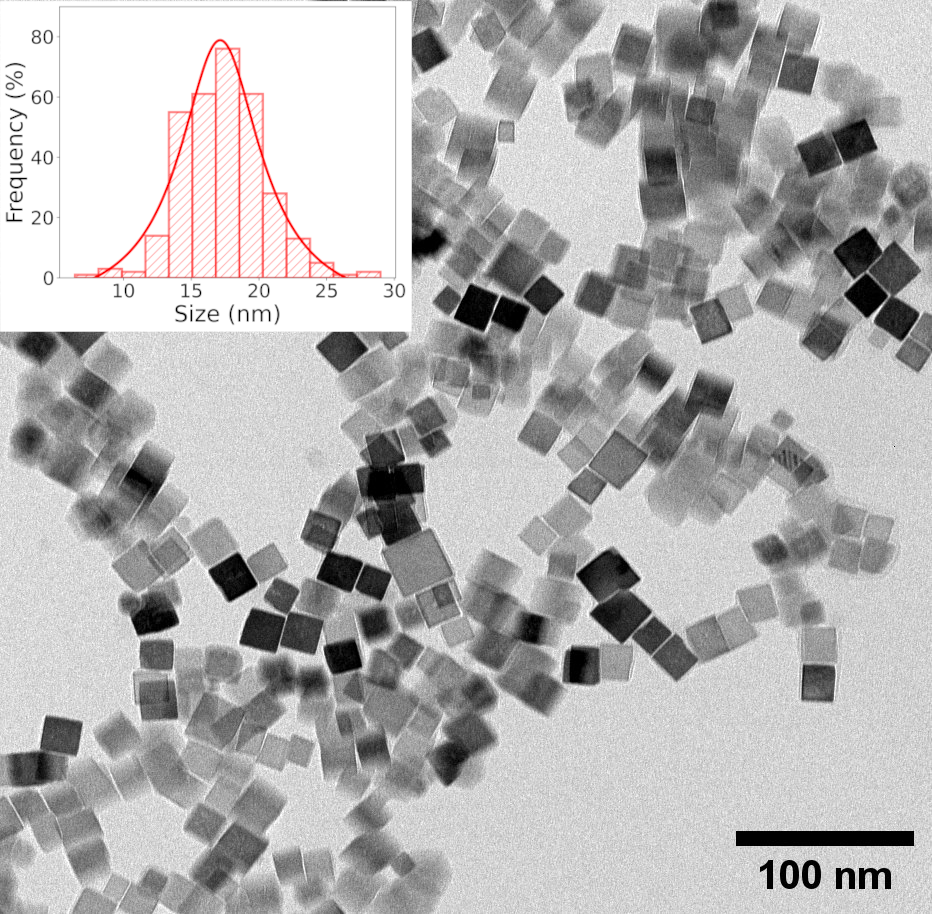}}

\caption{a) XRD patterns for 10~nm (bottom) and 20~nm (middle) and 400~nm (top) Co$_3$O$_4$ cubes. Red circles are the data, black solid line is the fit, and blue lines are the residuals of the refinement. The black Bragg markers index the Fd$\overline{3}$m structure.  Representative TEM images and size distributions from image analysis described in the text of the b) $\sim$10~nm and c) $\sim$20~nm nanocubes.\label{fig:TEM}}

\end{figure}


Figure~\ref{fig:XRD} presents the XRD patterns for the two nanocube and the larger cube samples.  All patterns are consistent with a single phase spinel structure (Fd$\overline{3}$m space group), and refinements using this space group resulted in lattice constants of 8.092~$\pm$~0.001~\AA~and 8.090$\pm$0.001~\AA~ for the nanocubes, and Scherrer broadening (from finite size effects) incorporated into the refinements identified $10\pm1$~nm and $17\pm~3$~nm cube sizes (using the necessary size parameter for a  cube shape\cite{Rabiei.2020} following the TEM image results (below).  The larger cubes have a lattice constant of 8.068$\pm$0.001~\AA, in agreement with the bulk value\cite{Roth.1964}.  The $\sim$0.3\% larger lattice constants of the nanocubes is a symptom of compression from the large fraction of surface atoms that create an unbalanced inward force that is offset by a core expansion, and a signature of a larger presence of intrinsic defects like oxygen vacancies that also act to expand the overall lattice.

TEM images and resultant size distributions are presented in figures~\ref{fig:TEM_10nm} and \ref{fig:TEM_20nm}.  Cube shapes are present throughout, and the fitted size distributions from \textgreater~500 nanoparticles (from a series of images sampling the TEM grid) identify 13$\pm$2~nm and 17$\pm$3~nm cube lengths from fits (solid lines in the figure insets), which we will be referred to as 10~nm and 20~nm sided cubes for simplicity.  The cubes (Supporting Information figure~1) have an average size of 380$\pm$50~nm, and we will refer to these cubes as 400~nm-sided.

\subsection{Antiferromagnetism?}


\begin{figure*}

\centering

\subfloat[a)\label{fig:invchi_vs_T_400nm}\label{fig:invchi_vs_T_nano}]{\includegraphics[width=0.325\textwidth]{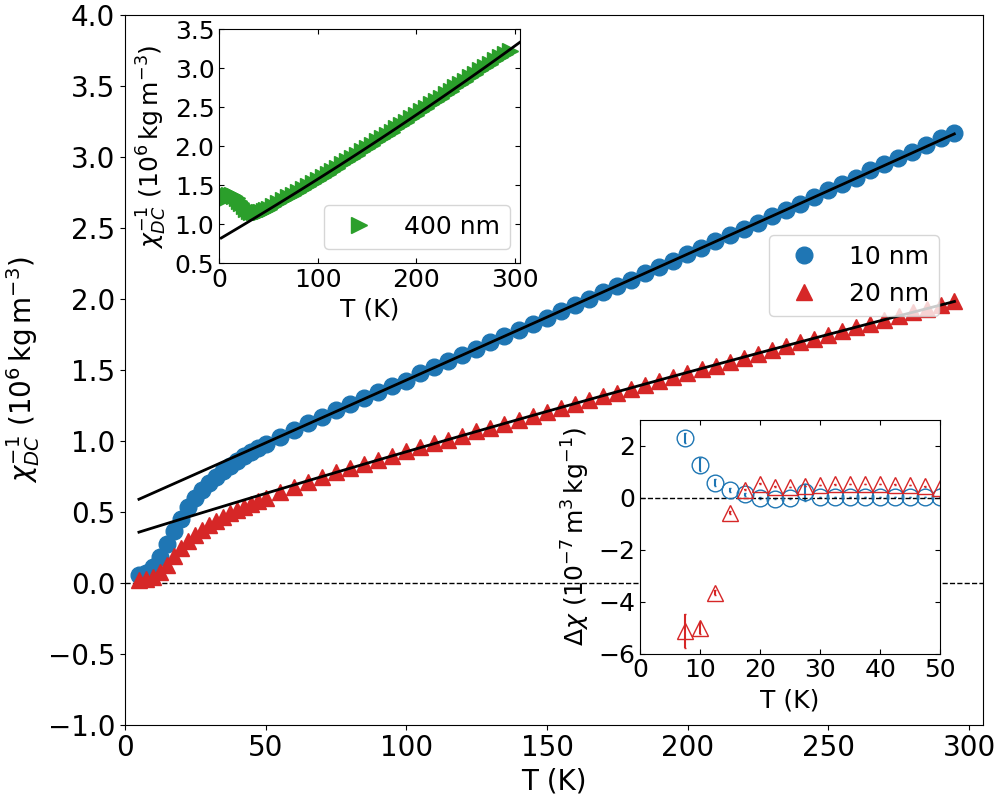}}
\hfill
\subfloat[b)\label{fig:MvsH_10nm}]{\includegraphics[width=0.325\textwidth]{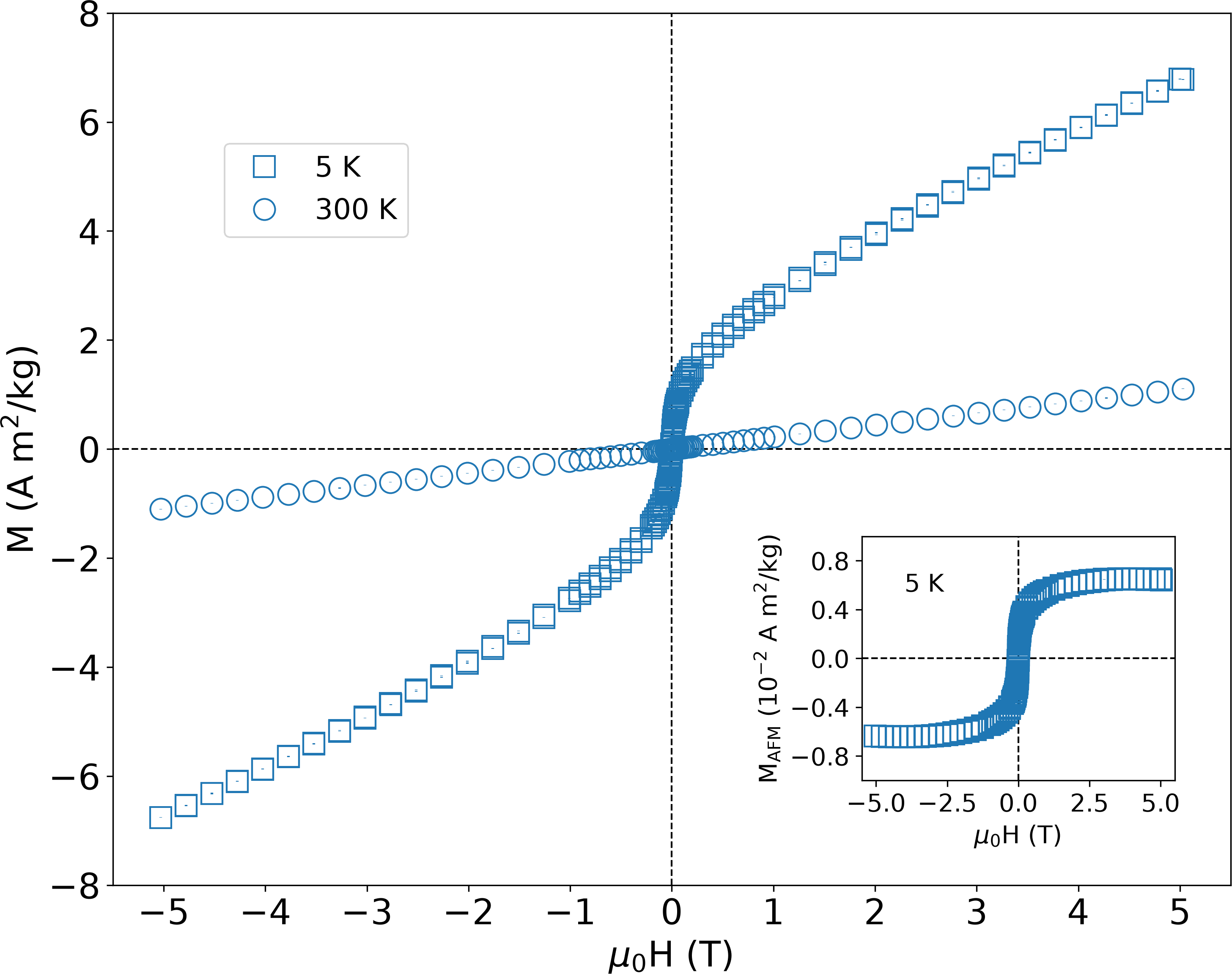}}
\hfill
\subfloat[c)\label{fig:MAFM_vs_T}]{\includegraphics[width=0.325\textwidth]{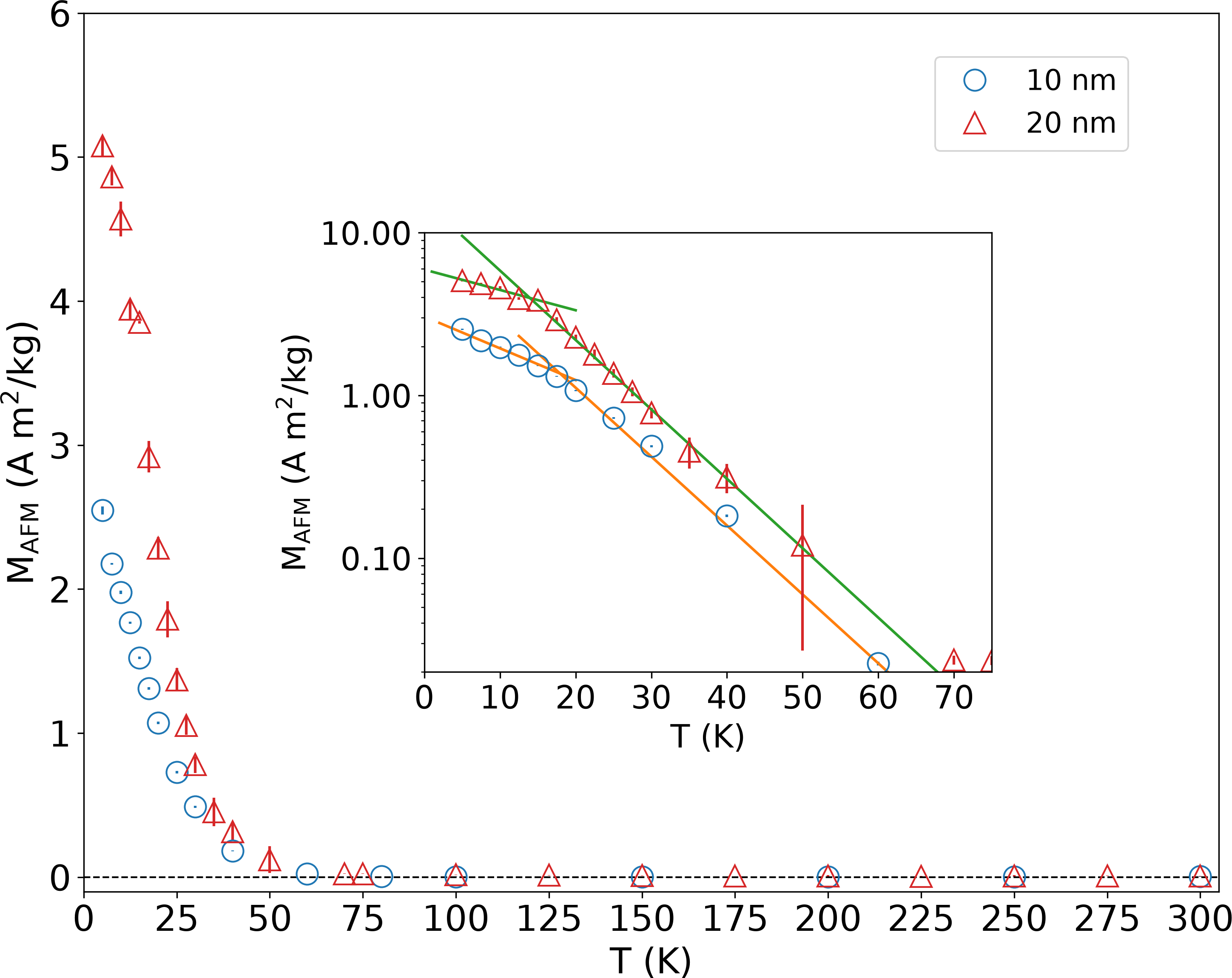}}

\caption{a) The temperature dependence of the inverse susceptibilities ($1/\chi_{\text{DC}}(T)$) of the two nanocube system.  The solid lines are fits described in the text.  Error bars are smaller than the symbol sizes.  The top inset shows $1/\chi_{\text{DC}}(T)$ of the 400~nm cubes, with the solid line a fit described in the text.  The bottom inset shows the temperature dependence of the difference between FC and ZFC $\chi(T)$, $\Delta \chi(T)$, for the 10 and 20~nm cubes.  b) Representative $M$ vs $\mu_0 H$ data at 5 and 300~K for the 10~nm nanocubes. The inset shows the $M_{\text{AFM}}$ component (described in the text) at 5~K.  c) Temperature dependence of the surface-spin component magnetization, $M_{\text{AFM}}(T)$, where the inset shows the exponential low temperature behaviour with the solid lines fits described in the text.}

\end{figure*}


The 400~nm cube's low field DC susceptibility ($\chi_{\text{DC}} = M/\mu_0 H$) temperature dependence (SI figure~S2) presents antiferromagnetic behaviour typical of bulk Co$_3$O$_4$ with a N\'eel temperature ($T_N$; marked by the maximum in $\chi_{\text{DC}}(T)$, and the inflection in $1/\chi_{\text{DC}}(T)$ in figure~\ref{fig:invchi_vs_T_400nm} between 35 and 40~K\cite{Roth.1964, Gawali.2018}.  In addition, the (small) irreversiblility in magnetization identified by the difference in field cooled (FC) and zero field-cooled (ZFC) $\chi_{\text{DC}}(T)$ that starts at $\sim$40~K (SI figure S2) denotes on the onset of field-aligned (canted) spins that can only occur below $T_N$.  The $T<30$~K $\chi_{\text{DC}}(T)$ decrease followed by an increase with cooling is a signature of the Co$^{2+}$ Kramers ions undergoing spin wave excitations\cite{Hadjiev.1988}.  Using a mean-field description of the overall magnetism, we fit the data using $1/\chi_{\text{DC}}(T)$ for $T>T_N$ with the Curie-Weiss law ($\chi_{\text{DC}}(T) = C/(T - \theta_W)$), where the Curie constant is $C = \mu_0 N \mu_{\text{eff}}^2 /k_B$, with $N$ the number of magnetic ions with moment $\mu$, $k_B$ Boltzmann's constant, and $\theta_W$ the Weiss temperature.  $T_N = (\{J(J+1)\}/\{3k_B\})(z_1 J_{1ex} - z_2 J_{2ex})$ and $\theta_W = (\{J(J+1)\}/\{3k_B\})(z_1 J_{1ex} + z_2 J_{2ex})$ where $J_{1ex}$ is the superexchange constant for the Co$^{2+}$(T$_{\text{d}}$)--O--Co$^{2+}$(T$_{\text{d}}$) pathway with $z_1$=4 nearest Co$^{2+}$ neighbours, and $J_{2ex}$ is the superexchange constant for the Co$^{2+}$(T$_{\text{d}}$)--O--Co$^{3+}$(O$_{\text{h}}$)--O--Co$^{2+}$(T$_{\text{d}}$) pathway with $z_2$=12 nearest neighbours.  We find (solid line in figure.~\ref{fig:invchi_vs_T_400nm}) that $C=1.6\pm0.1 \times 10^{-4}$~Km$^3$/kg, $\theta_W=-130\pm2$~K, and with $\mu_{\text{eff}}=5.03\pm0.04$~$\mu_B$, have $J_{1ex}=9.3\pm0.1$~K and $J_{2ex}=2.06\pm0.01$~K.  Within the limitations of this mean-field description, these $\mu_{\text{eff}}$ and $J_{ex}$ values are within 15\% of the bulk \cite{Dutta.2008} and in good agreement. As expected, $\mu_{\text{eff}}$ is higher than the spin-only $S=3/2$ 3.87~$\mu_B$ for Co$^{2+}$, and is likely the results of a recaptured orbital component of the ions' $^4$F$_{9/2}$ ground state\cite{Morrish.book, Coey.book}.

By contrast, the 10~nm and 20~nm cubes' low field DC susceptibilities temperature dependence, shown in figure~\ref{fig:invchi_vs_T_nano} and SI figure~S3 reveals quite different nanomagnetism.  Not only is the signature cusp of a $T_N$ not present (indicative of significant suppression of magnetic ordering due to much weaker $J_{ex}$), but the maximum $\chi_{\text{DC}}(T)$ values are $\sim$200 (10~nm cubes) to $\sim$500 (20~nm cubes) times larger.  This significant enhancement of magnetic response is indicative of more and larger moment Co spins responding to the applied field.  Like the 400~nm cubes, we see a divergence between the FC and ZFC $\chi_{\text{DC}}(T)$ ($\Delta \chi_{\text{DC}}(T)$, inset in figure~\ref{fig:invchi_vs_T_nano}) that denotes a $T_N\sim12.5$~K for the 10~nm nanocubes and $T_N\sim10$~K for the 20~nm nanocubes.  These $T_N$s were corroborated using Fisher analysis\cite{Fisher.1962} (not shown).  To quantify the apparent decrease in super-exchange constants due to finite-size effects, we turn to a similar Curie-Weiss anlaysis as discussed above, but include a temperature independent (diamagnetic) susceptibility term, $\chi_0$, to capture to complete high temperature behaviour\cite{Dutta.2008}, i.e. $\chi_{\text{DC}}(T) = \chi_0 + C/(T - \theta_W)$.  Fits are shown by the solid lines in figure~\ref{fig:invchi_vs_T_nano}.  We find $\chi_0 = -0.4\pm0.3 \times 10^{-8}$~m$^3$/kg for the 10~nm nanocubes and $\chi_0 = -5.3\pm0.4 \times 10^{-8}$~m$^3$/kg for the 20~nm nanocubes, in good agreement with the bulk value of $\sim -3\times 10^{-8}$~m$^3$/kg\cite{Dutta.2008}.  $C=1.1\pm0.1 \times 10^{-4} $~Km$^3$/kg and $1.5\pm0.1 \times 10 ^{-4}$~Km$^3$/kg for the 10 and 20~nm nanocubes, respectively, while $\theta_W = -52\pm1$~K and $-62\pm1$~K for the 10 and 20~nm nanocubes.  $C$ and $T_N$ values identify $\mu_{\text{eff}}=4.90\pm0.03~\mu_B$ for the 10~nm nanocubes, and $\mu_{\text{eff}}=4.18\pm0.02~\mu_B$ for the 20~nm nanocubes.  These $\mu_{\text{eff}}$ values are lower than the bulk-like 400~nm cubes $\mu_{\text{eff}}$, so the two-orders of magnitude increase in $\chi_{\text{DC}}(T)$ below $T_N$ for the nanocubes is unlikely to be from a further enhancement of the Co$^{2+}$ orbital moment contribution discussed above\cite{Dutta.2008,Coey.book}.  Indeed, the $\sim$15\% decrease in $\mu_{\text{eff}}$ with the factor of $\sim8$ increase in volume between the 10 and 20~nm nanocubes hints at a more complex interplay between Co$^{2+}$ and Co$^{3+}$ spin moment behaviours (e.g. Co$^{3+}$ should be in a low spin state in Co$_3$O$_4$ and have no moment) than implied by `simple' finite-size effects\cite{Coey.2019, Paidi.2019, Batlle.2002}.  A further suggestion to a more subtle interplay between electronic and magnetic configurations of the Co ions is that the superexchange constant $J_{1ex}$ decreases from $6.38\pm0.07$~K to $3.86\pm0.08$~K and $J_{2ex}$ decreases from $1.41\pm0.01$~K to $0.87\pm0.01$~K for the 10 and 20~nm nanocubes, respectively.  All the above results suggest strongly that the Co$^{2+}$(T$_{\text{d}}$)--O--Co$^{2+}$(T$_{\text{d}}$) and Co$^{2+}$(T$_{\text{d}}$)--O--Co$^{3+}$(O$_{\text{h}}$)--O--Co$^{2+}$(T$_{\text{d}}$)) superexchange routes responsible for the nanomagnetism have been altered significantly, and not just by the relative changes of altered and broken surface coordination.


\begin{figure*}

\centering

\subfloat[a) \label{fig:XANES}]{\includegraphics[width=0.25\textwidth]{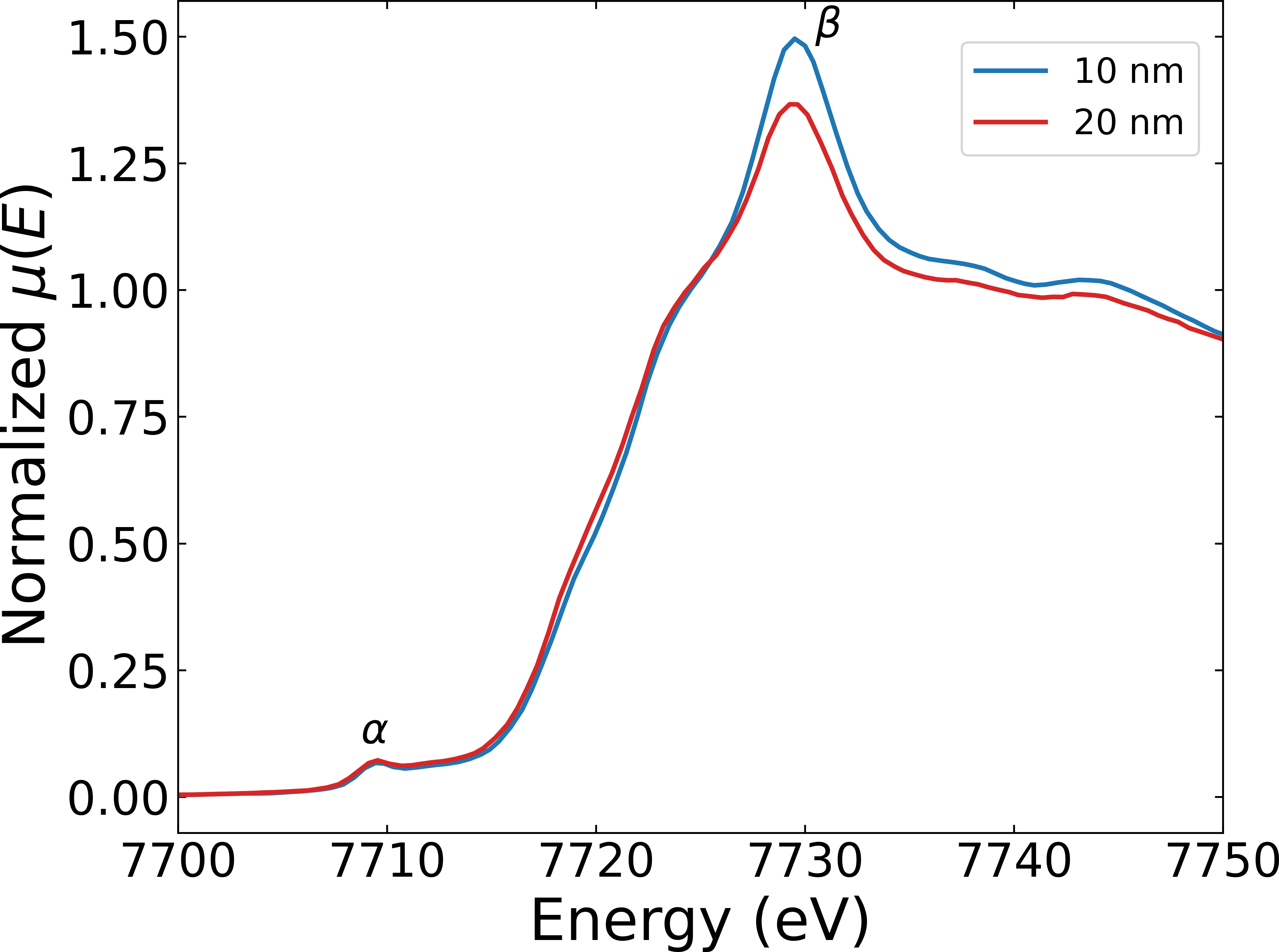}}
\hfill
\subfloat[b) \label{fig:EXAFS_10nm}]{\includegraphics[width=0.35\textwidth]{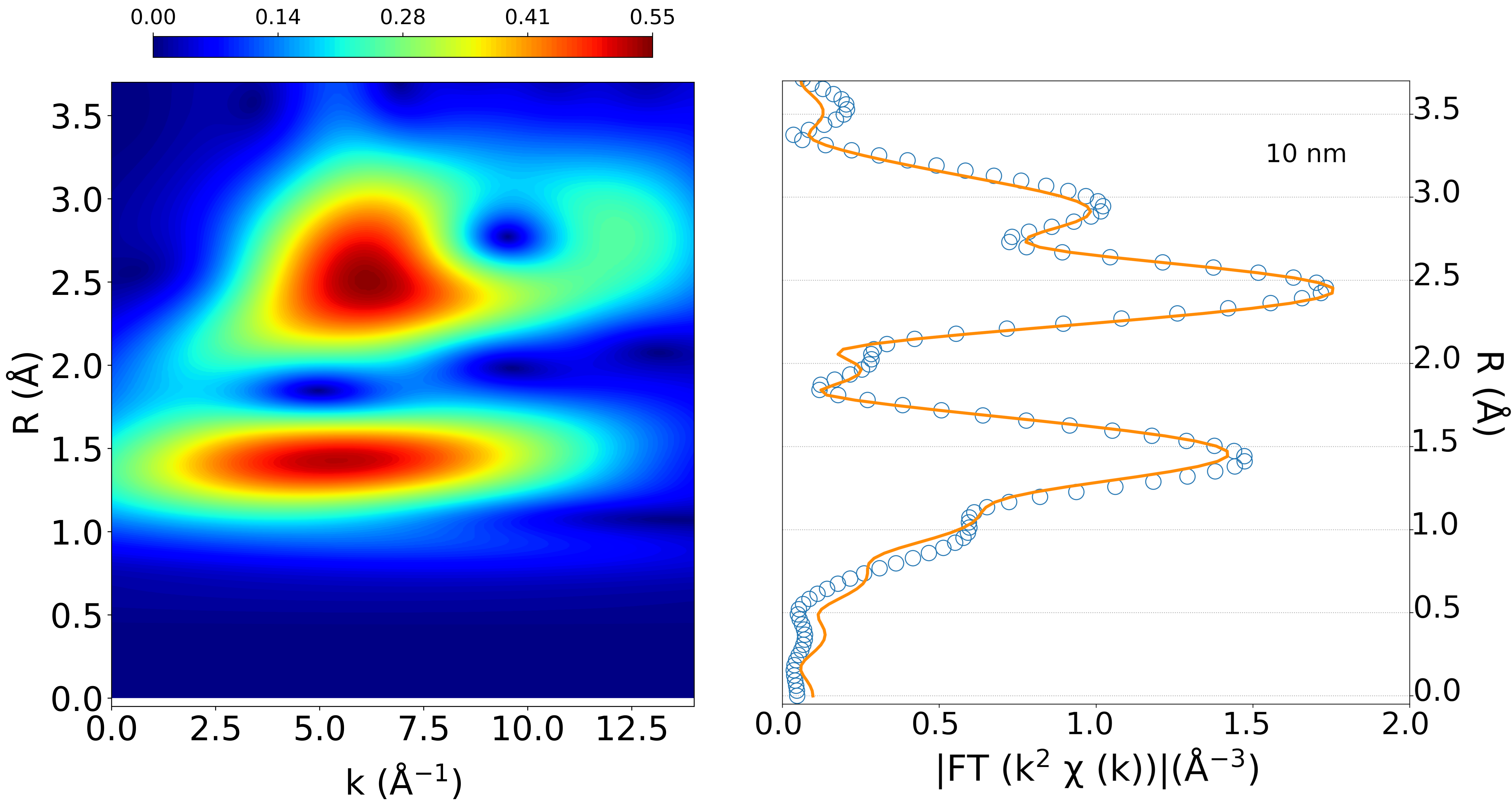}}
\hfill
\subfloat[c) \label{fig:EXAFS_20nm}]{\includegraphics[width=0.35\textwidth]{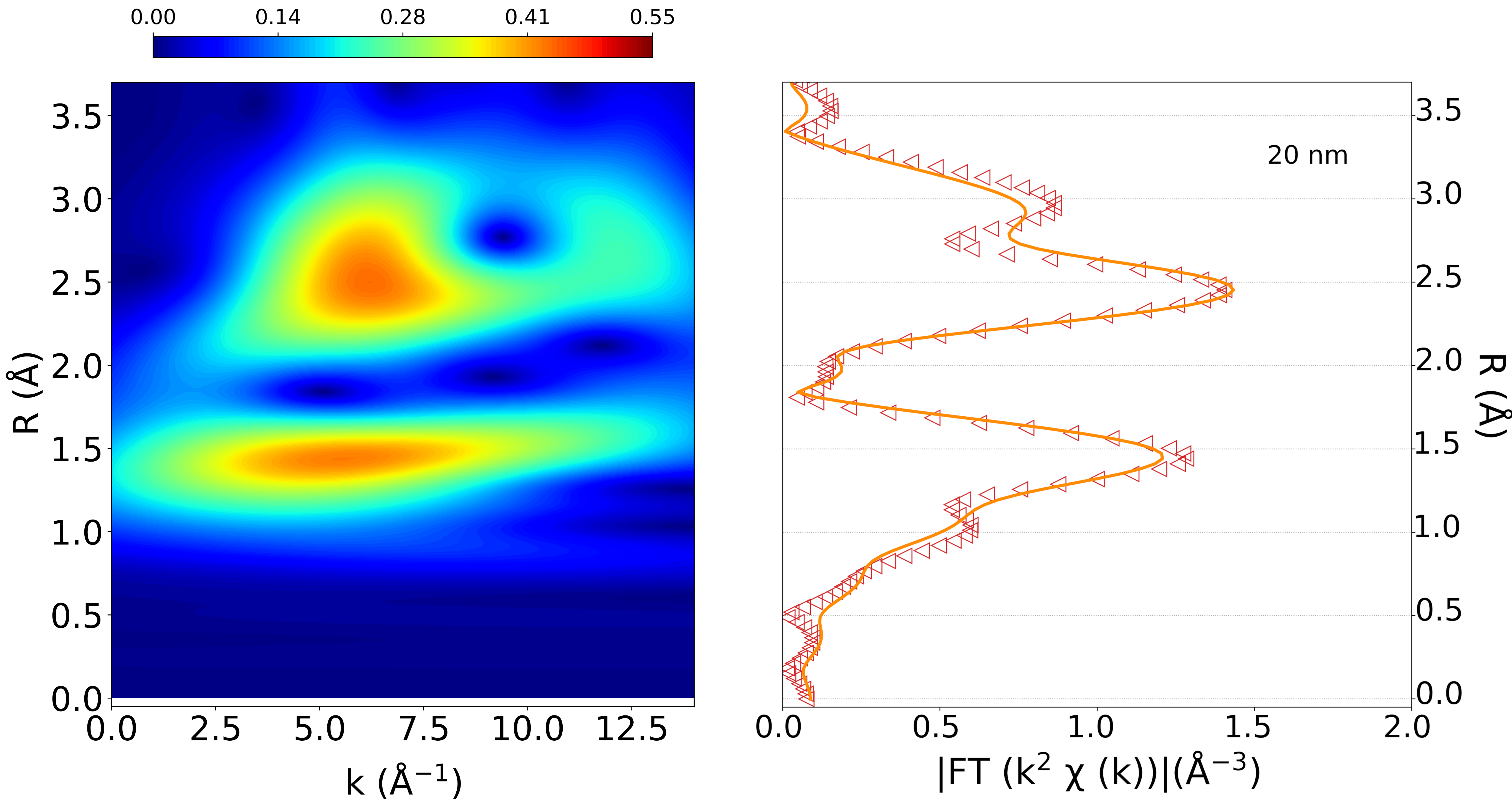}}

\subfloat[d) \label{fig:Co_XAS_10nm}]{\includegraphics[width=0.325\textwidth]{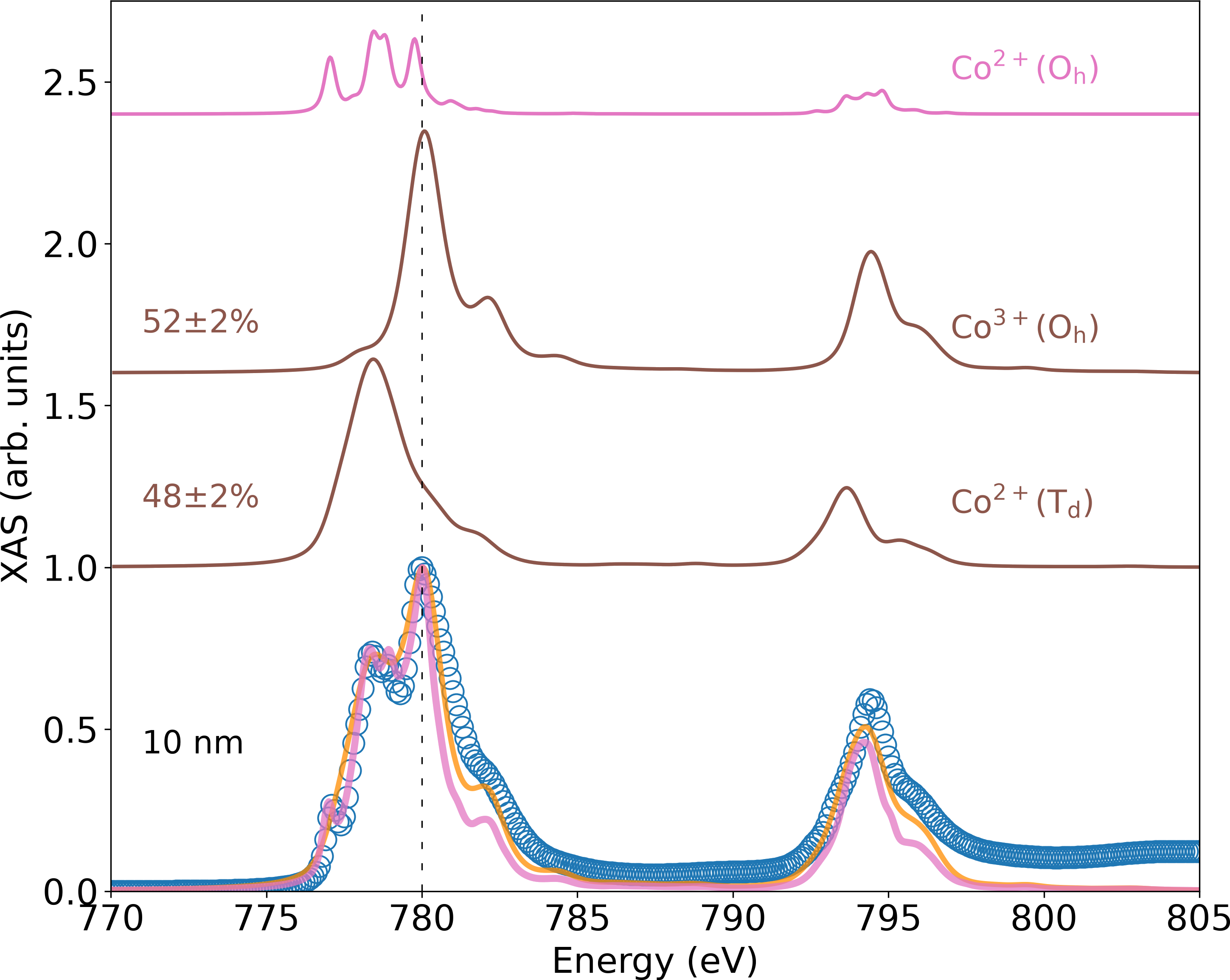}}
\hfill
\subfloat[e) \label{fig:Co_XAS_20nm}]{\includegraphics[width=0.325\textwidth]{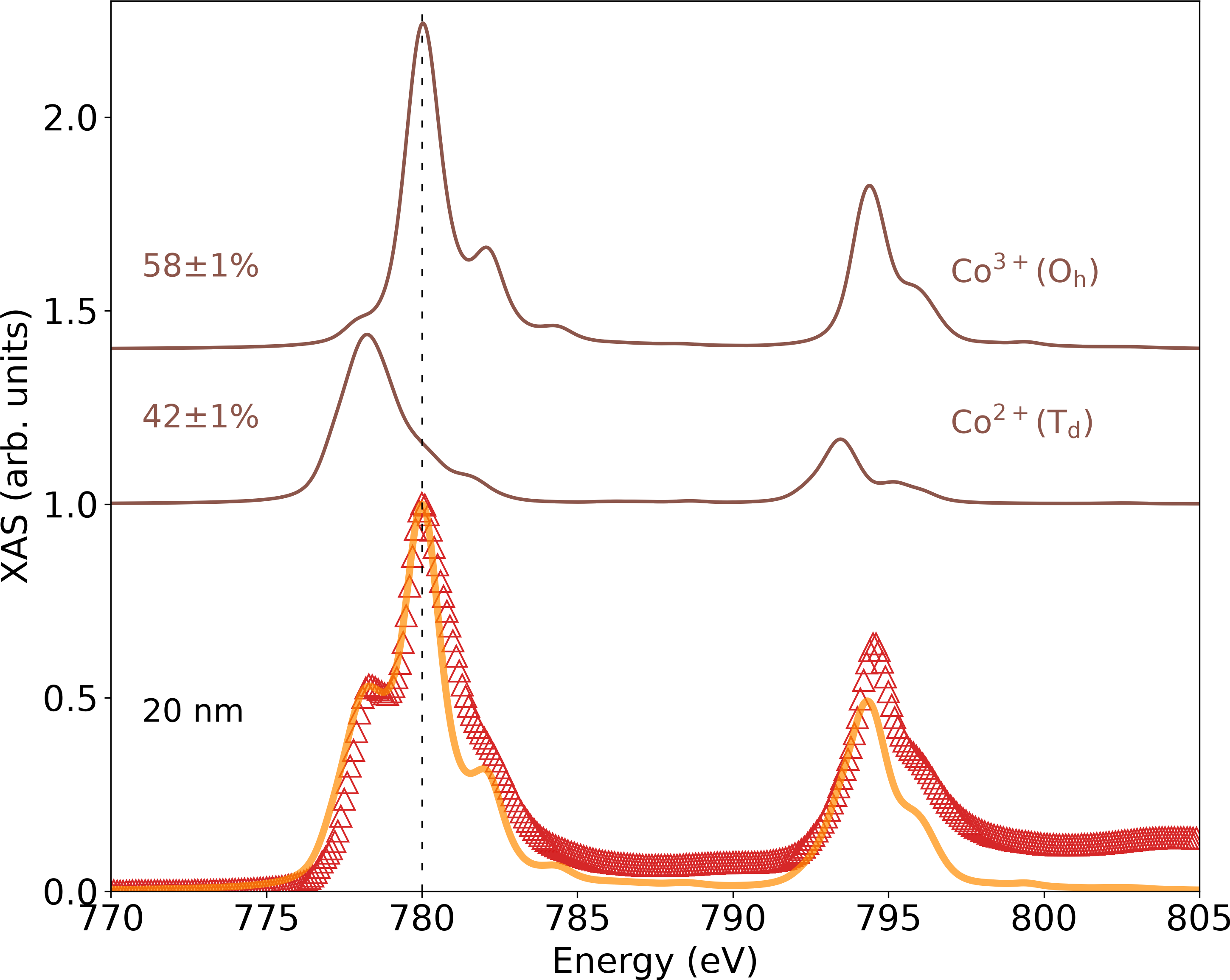}}
\hfill
\subfloat[f) \label{fig:O_XAS}]{\includegraphics[width=0.325\textwidth]{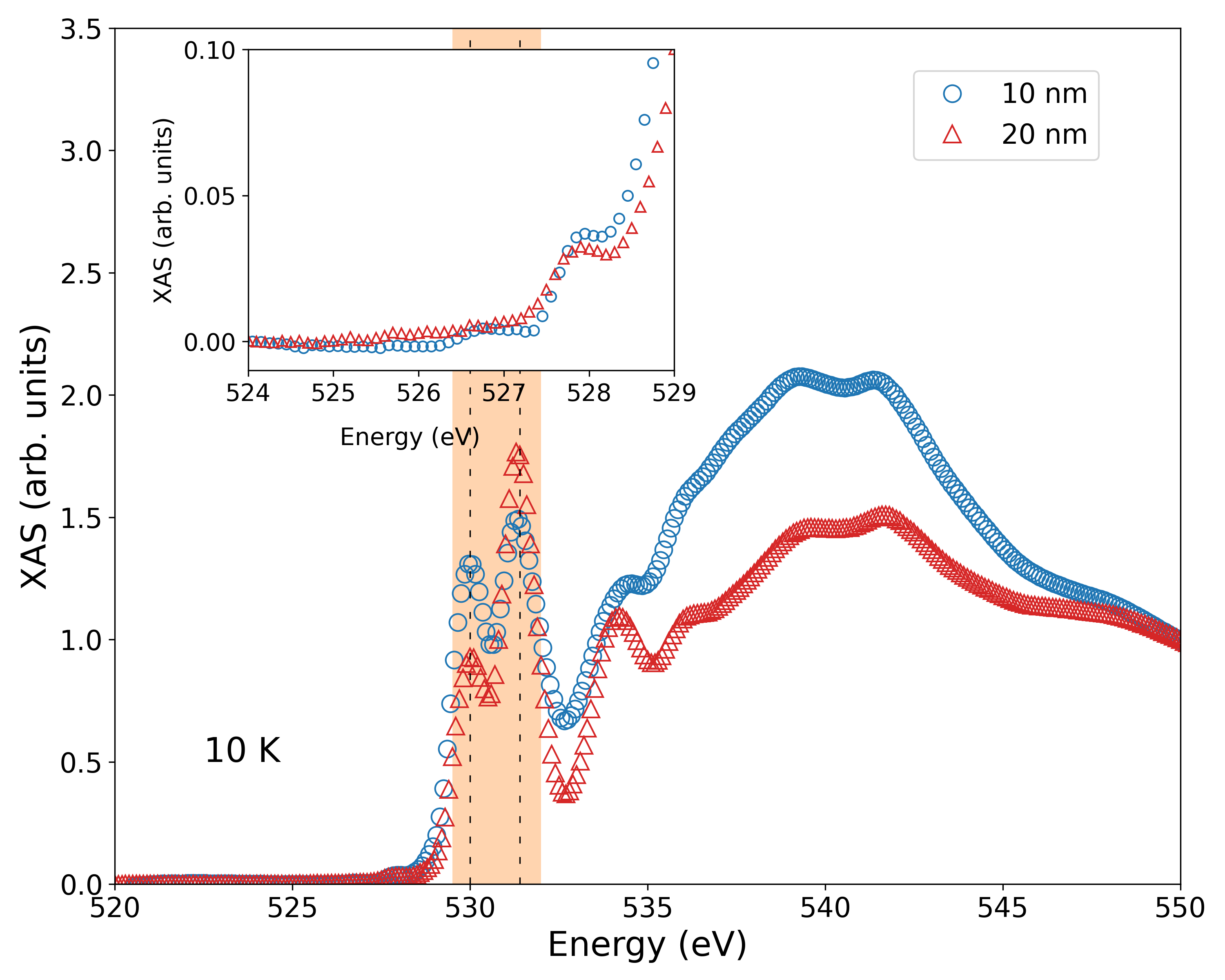}}

\subfloat[g) \label{fig:Co_XMCD_10nm_10K}]{\includegraphics[width=0.325\textwidth]{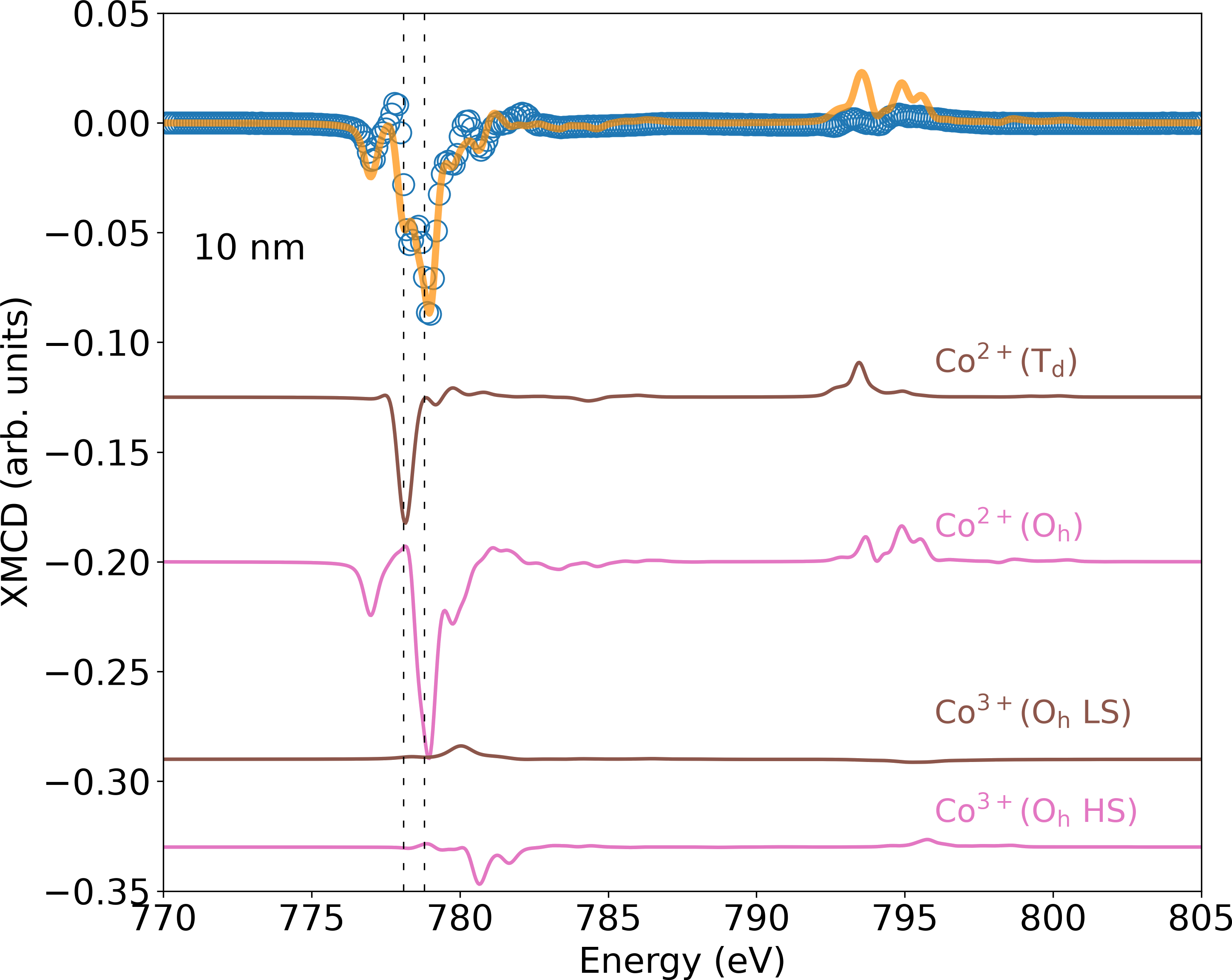}}
\hfill
\subfloat[h) \label{fig:Co_XMCD_20nm_10K}]{\includegraphics[width=0.325\textwidth]{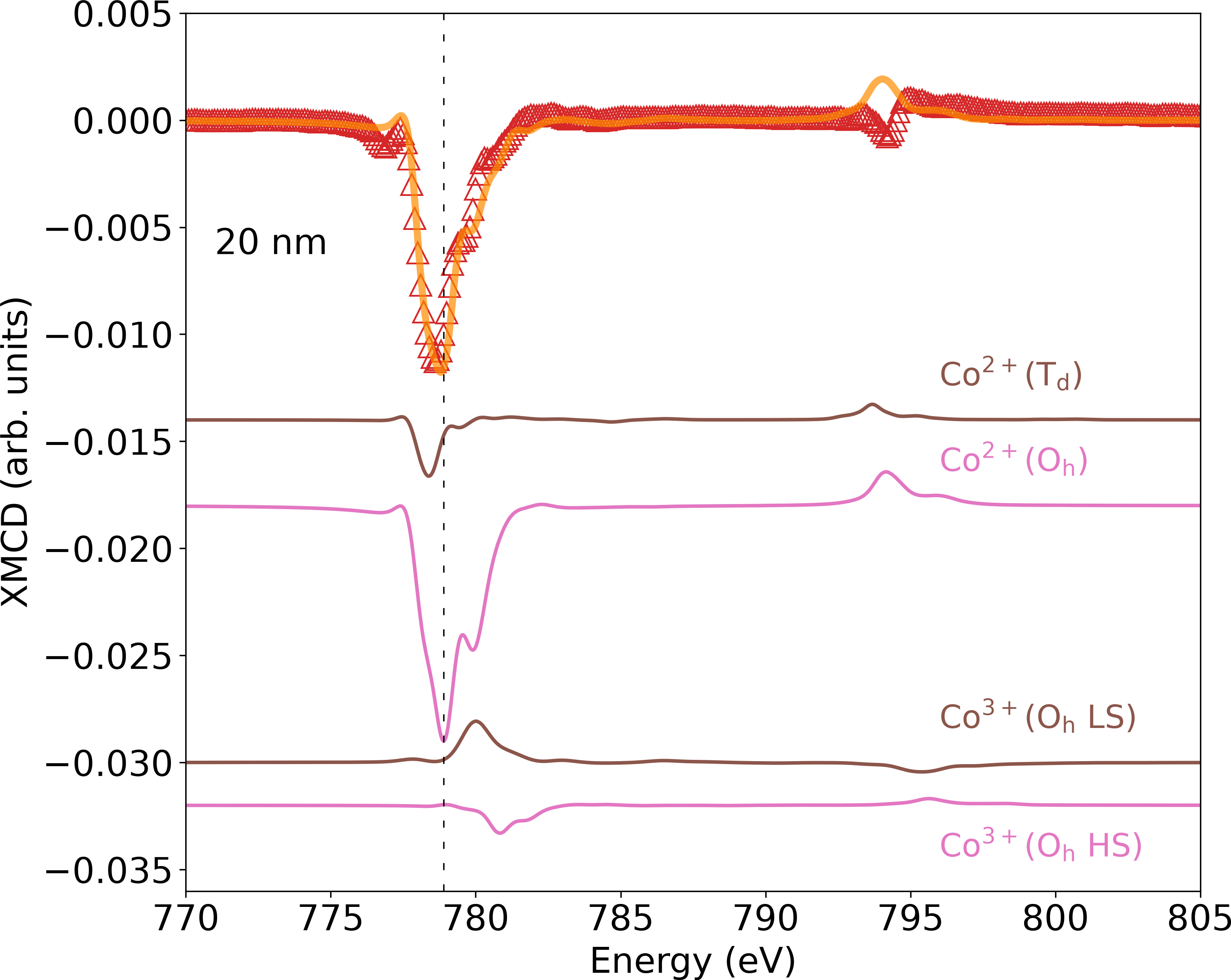}}
\hfill
\subfloat[i) \label{fig:M_Co_XMCD}]{\includegraphics[width=0.325\textwidth]{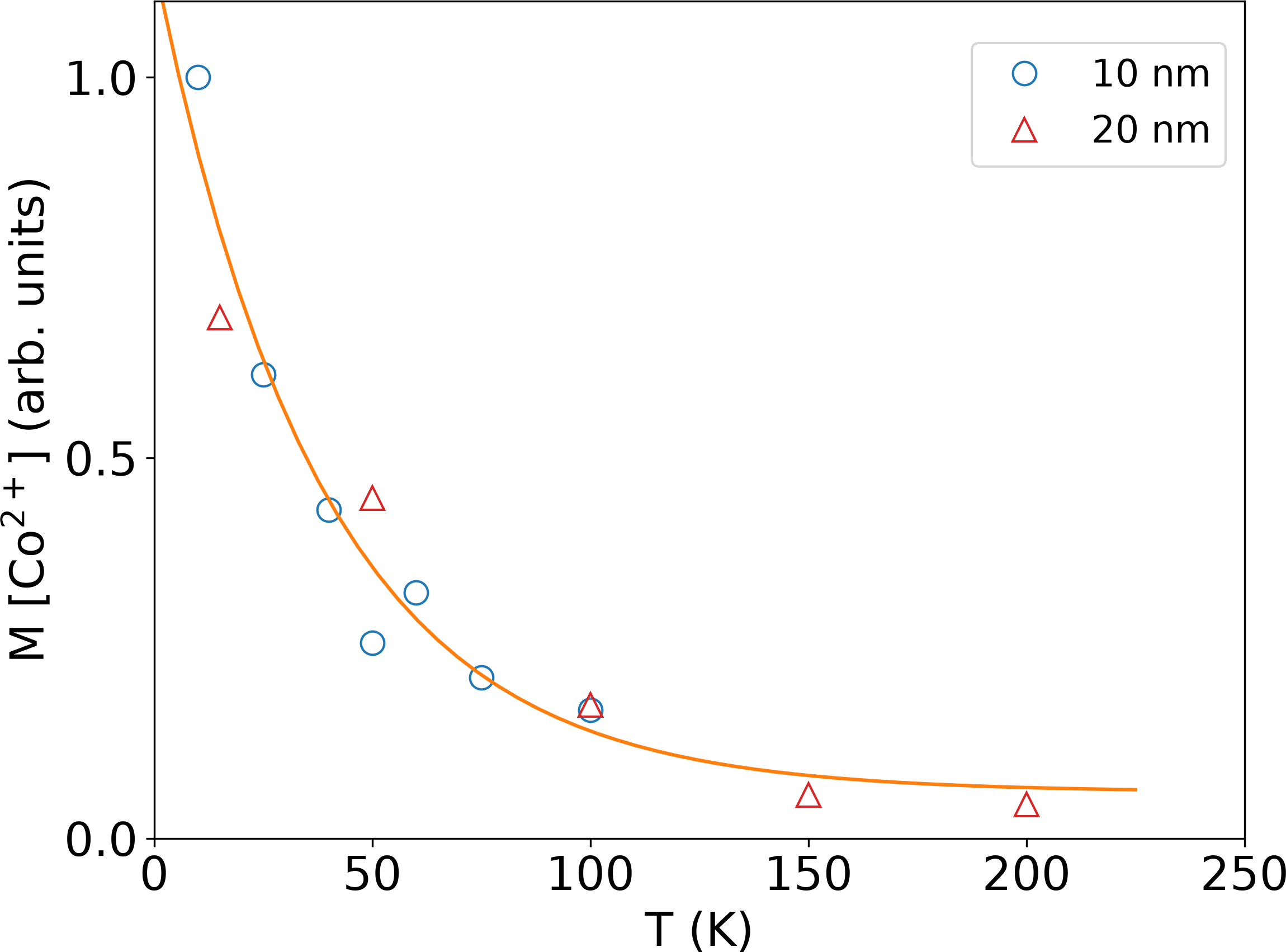}}

\caption{a) Normalized Co K-edge XANES spectra for the 10 and 20~nm Co$_3$O$_4$ nanocubes and 300~K, and b \& c) the corresponding EXAFS spectra in real space along with respective wavelet transforms.  The fits (described in the text) are represented by the solid lines through the data.  The peaks at $\sim$1.45~\AA~are from Co-O scattering, while the peaks at 2.5~\AA~and 3.0~\AA~in the Co K edge EXAFS are primarily due to the scattering paths of Co(O$_{\text{h}}$)-Co(O$_{\text{h}}$) and Co(O$_{\text{h}}$/T$_{\text{d}}$)-Co(T$_{\text{d}}$), respectively. \label{fig:XAFS}  Normalized Co L$_{\text{3,2}}$-edge XAS spectra at 10~K for the d) 10 and e) 20~nm Co$_3$O$_4$ nanocubes.  The ligand-field multiplet simulation for each Co coordination in the spinel structure (scaled by the relative fraction) is presented by the brown lines, with the fit of these components the orange line through the data.  The pink solid like is the unusual Co$^{2+}$ O$_{\text{h}}$ coordination (not scaled), with the pink dashed line through the data a fit with that additional component.  All simulations are described in detail in the text.\label{fig:Co_XAS} f) Normalized O K-edge XAS spectra at 10~K for the 10 (blue $\bigcirc$) and b) 20~nm (red $\bigtriangleup$) Co$_3$O$_4$ nanocubes at 10~K.  The inset shows the pre-edge region.  Co L$_{\text{3,2}}$-edge XMCD spectra at 10~K for the g) 10 and h) 20~nm Co$_3$O$_4$ nanocubes.  The ligand-field multiplet simulation for each Co coordination in the spinel structure is presented by the brown (usual coordination) and pink (unusual coordination) lines, with the fit of these components the orange line through the data.  All simulations are described in detail in the text.  i) Temperature dependence of the principal Co$^{2+}$ elemental magnetization of the Co$_3$O$_4$ nanocubes of the 10 and 20~nm Co$_3$O$_4$ nanocubes, normalized with respect to M$_{\text{AFM}}$(5~K). The line is a guide to the eye.}

\end{figure*}


The low-field $\chi_{\text{DC}}(T)$ measures the overall magnetization response of the Co$_3$O$_4$ nanocubes.  It is anticipated that there will be surface Co$^{2+}$ ions that suffer compromised coordination (e.g. missing neighbours and dangling bonds) that yield (surface) spins that present an overall measured magnetization.  This surface spin contribution should also enable an exchange bias effect due to an interfacial spin torque between uncompensated surface and antiferromagnetic core spins.  To pinpoint if the surface spins are responsible for the trends identified by $\chi_{\text{DC}}(T)$, we turn to the field-dependence of the magnetization ($M$ vs $\mu_0H$) of the Co$_3$O$_4$ nanocubes using $\pm5$~T after 5~T field cooling from 300~K, shown in figure~\ref{fig:MvsH_10nm}.  The 5~T field should affect all the spins that are not fully compensated via antiferromagnetic exchange, and thus permit a separation of surface and core nanomagnetism by revealing a surface magnetic anisotropy (measured coercivity, $\mu_0 H$), and a core (antiferromagnetic) contribution.  Figure~\ref{fig:MvsH_10nm} presents representative data for the 10~nm Co$_3$O$_4$ nanocubes, and the 20~nm Co$_3$O$_4$ nanocubes exhibited essentially the same $M$ vs $\mu_0H$ behaviour over the same temperature range (5 to 300~K).  Firstly, no coercivity ($\mu_0 H_c(T) = 0$) was measured, and hence, no exchange bias, indicating the lack of a surface anisotopy and no real magnetic order among any free surface spins on the nanocubes. The field-dependence of the magnetization follows the usual prescription of $M(\mu_0 H) = M_{\text{AFM}}(\mu_0 H) + \chi_{\text{HF}}(\mu_0 H)$, where $M_{\text{AFM}}$ captures the N\'eel uncompensated magnetization of surface spins\cite{Neel.1961, Mollard.1973} and $\chi_{\text{HF}}$ describes the canting of the Co$_3$O$_4$ nanocubes' core antiferromagnetic sublattice with the applied field\cite{Krasikov.2023}.  For the Co$_3$O$_4$ nanocubes, $\chi_{\text{HF}}(T)$ was determined from linear fits to the $M(\mu_0 H \textgreater 1~T)(T)$ data, with the fit results presented in SI figure~S3.  Both 10~nm and 20~nm high-field $\chi_{\text{tot}}(T)$ have the same overall behaviour measured by the low-field $\chi_{\text{DC}}(T)$ in figure~\ref{fig:invchi_vs_T_nano}, clearly linking this nanomagnetism to the nanocubes' core.  Indeed, the same analysis as discussed above identified similar $T_N$ and $\mu_{\text{eff}}$, and for the 10~nm Co$_3$O$_4$ nanocubes we find $J_{1ex}=6.5\pm0.02$~K and $J_{2ex}=1.52\pm0.02$~K, and $J_{1ex}=4.1\pm0.4$~K and $J_{2ex}=0.88\pm0.03$~K, values in excellent agreement with the low-field analysis.  Clearly, the reduced size of the nanocubes results in significantly weakened super-exchange interactions throughout the cube volumes and not just on their surfaces, unlike the majority of antiferromagnetic nanoparticles\cite{Liu.2025,Morup.2007}.

The uncompensated magnetization's temperature dependence, $M_{\text{AFM}}(T)$, from the nanocubes' surface spins, is shown in figure~\ref{fig:MAFM_vs_T}.  Like the low-field $\chi_{\text{DC}}(T)$ in figure~\ref{fig:invchi_vs_T_nano}, the 10~nm nanocubes' $M_{\text{AFM}}(T)$ is less than the 20~nm nanocubes' for $T\lesssim$50~K.  But, the fractional difference between 10~nm and 20~nm nanocubes $\chi_{\text{DC}}(T)$ and $M_{\text{AFM}}(T)$ is less than than expected from a simple geometrical scaling (e.g. surface area or volume), something more subtle and complex (e.g. oxygen vacancies and redistribution of Co coordinations and valences) is likely responsible.  These surface spins freeze significantly above the nanocubes' $T_N$s, where $M_{\text{AFM}}(T)$\textgreater 0 below $\sim$50~K, in the same temperature regime where short-range spin correlation behaviour is presented by the divergence from $1/\chi_{\text{DC}}(T)$ behaviour in figure~\ref{fig:invchi_vs_T_nano}.  This is all indicative of unconventional, likely short-range `ordering' process among the surface spins.  Closer inspection of $M_{\text{AFM}}(T)$ reveals an exponential-like temperature dependence (inset of figure~\ref{fig:MAFM_vs_T}) that is a hallmark of freezing surface spins upon cooling\cite{Mandal.2006, Cojocaru.2011, Alkadour.2016}.  The origin of this temperature dependence is likely from an energy gap in the spin wave-spectrum due to its quantization from finite size effects.  But, there is a clear break in $M_{\text{AFM}}(T)$ around $\sim20$~K for both 10 and 20~nm nanocube systems that hints at a break in the spin wave spectrum from the spatial disorder and dynamics, much like a spin-glass\cite{Moskalenko.1981, Tahriri.2027}.  To quantify all the above unusual behaviour, we need to delve into the local, atomic structure and coordination of the Co$_3$O$_4$ nanocubes, as well as what specifically the Co$^{2+}$ and Co$^{3+}$ are doing magnetically in their respective O$_{\text{h}}$ and T$_{\text{d}}$ sites.

\subsection{Radical changes to the electronic structure and disorder}

We first turn to Co K-edge spectroscopy (x-ray absorption fine structure, XAFS) at 300~K to characterize the bond distances and structure (and its disorder) throughout the 10 and 20~nm nanocubes.  Figure~\ref{fig:XAFS} presents the XAFS spectra of the Co ions over the x-ray absorption near edge structure (XANES) and extended x-ray absorption fine structure (EXAFS) regimes.  XANES (region $\alpha$ to just before $\beta$ in figure~\ref{fig:XANES}) is exquisitely sensitive to the valence of the Co ions and its coordination geometry, while EXAFS (region $\beta$ and beyond in figure~\ref{fig:XANES}) permits a quantification of the numbers of bonds, their types, and their bond distances of surrounding Co ions that the x-ray scattered from.  The 10 and 20~nm Co$_3$O$_4$ nanocubes' XANES is identical, showing that the average Co valences are size (surface) independent, and its spectral shape is consistent with bulk Co$_3$O$_4$\cite{Bai.2012, Mijiti.2021}.  Specifically, the XANES region is scattering from dipole allowed transitions from 1s to 4p orbitals that can be hybridized with the unoccupied 3d orbitals in the Co ions present in T$_{\text{d}}$ environments. The broader spectral features at the higher energies from $\sim \alpha$ to before the $\beta$ peak in figure~\ref{fig:XANES} corresponds to the quadrupole allowed transition from 1s to unoccupied 3d orbitals in the Co ions residing in the O$_{\text{h}}$ environments. Since there is no energy dependent change in these feature, the density of unoccupied 3d states do not change on increasing nanocube size from 10 to 20~nm.


\begin{figure*}

\centering

\subfloat[a) \label{fig:O_XAS_vs_T_10nm}]{\includegraphics[width=0.325\textwidth]{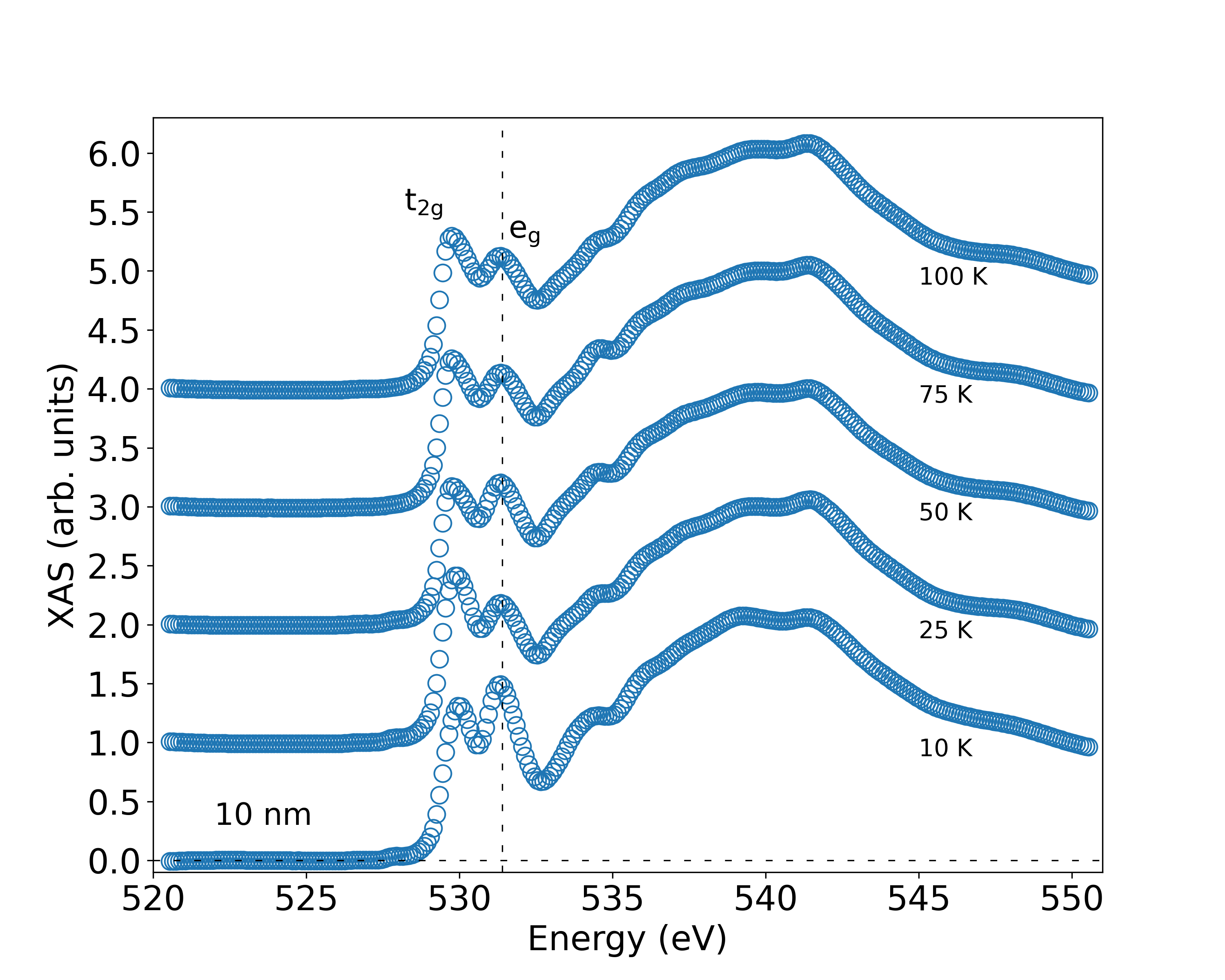}}
\hfill
\subfloat[b) \label{fig:O_XAS_vs_T_20nm}]{\includegraphics[width=0.325\textwidth]{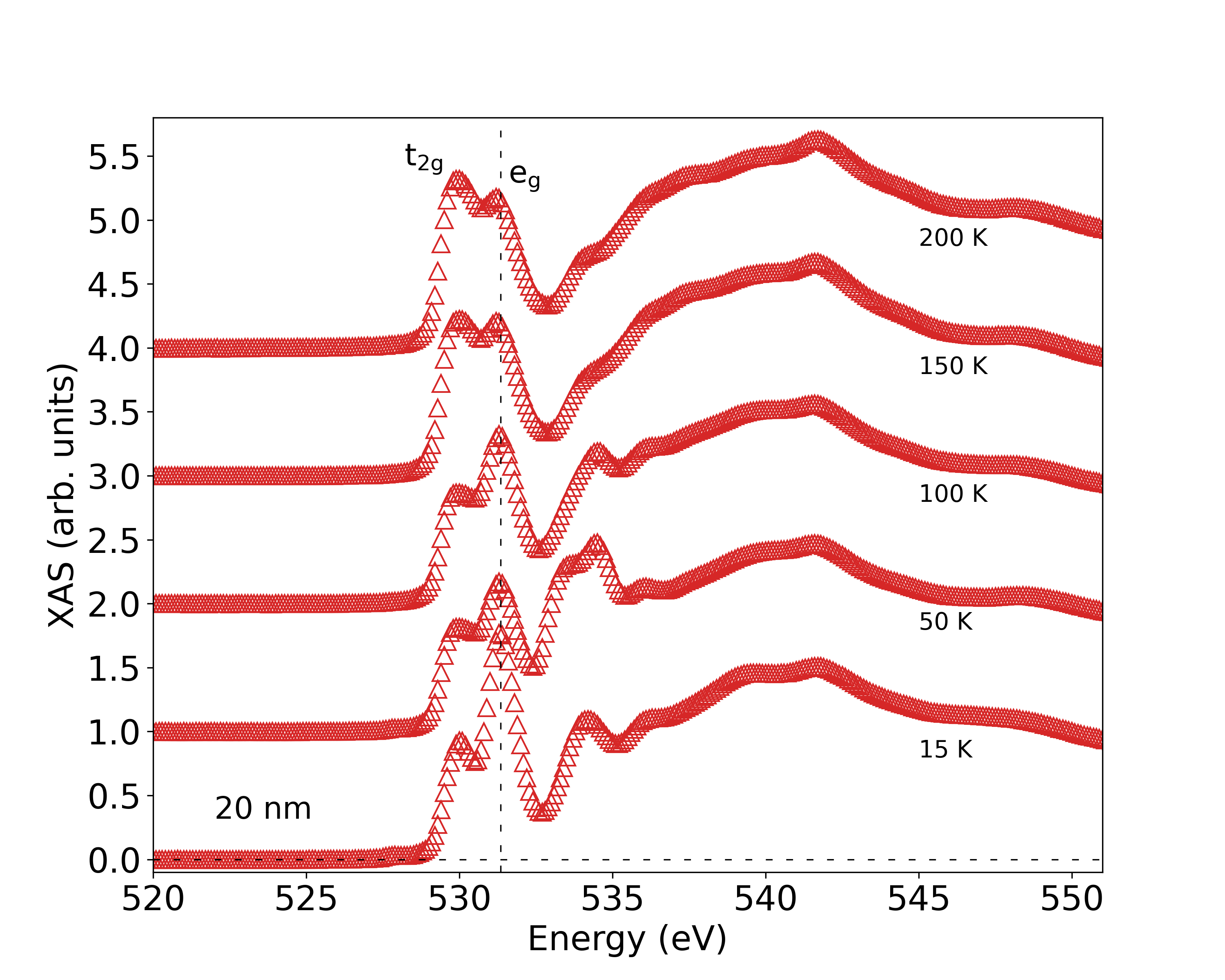}}
\hfill
\subfloat[c) \label{fig:t2geg_vs_T}]{\includegraphics[width=0.325\textwidth]{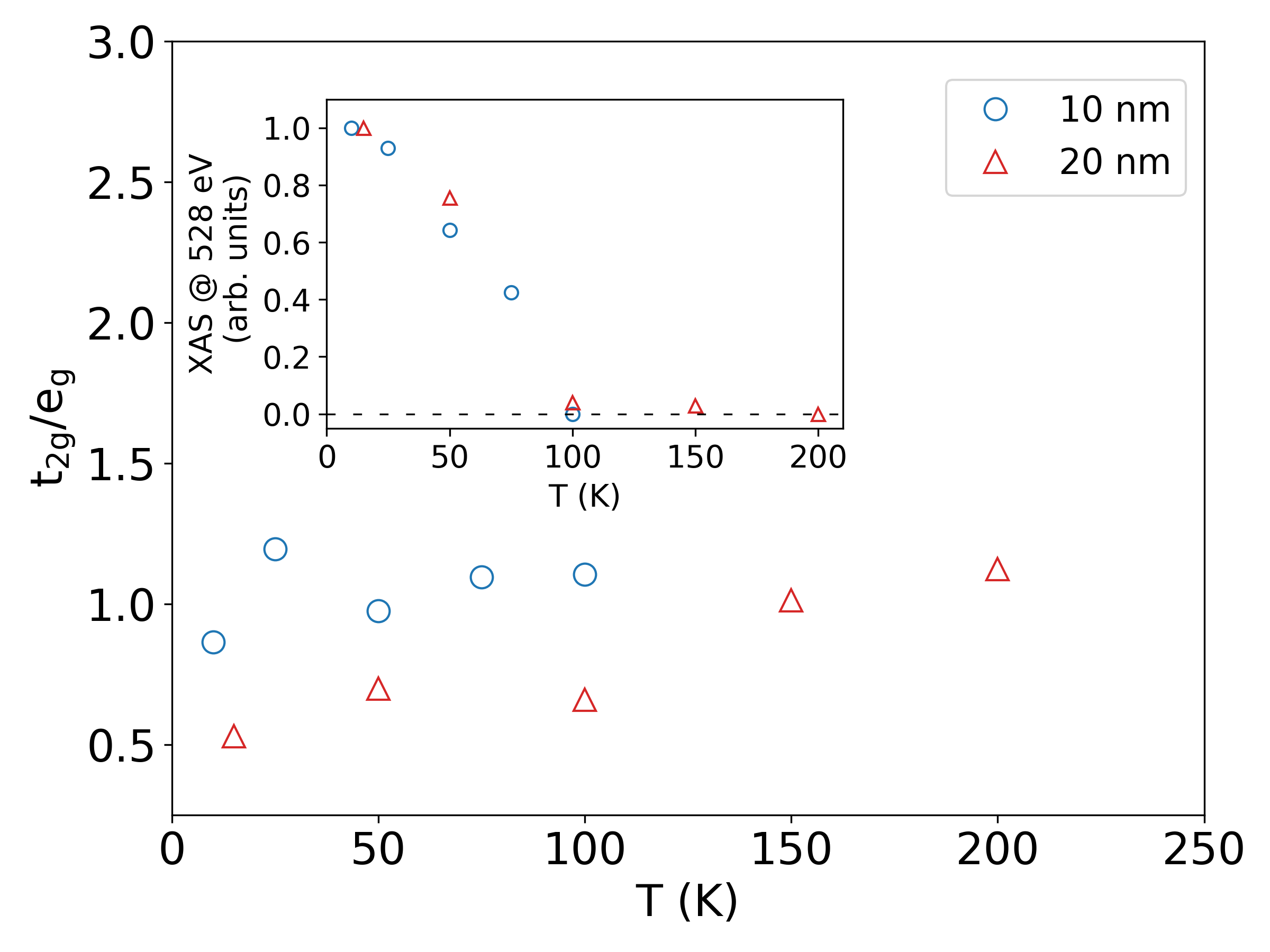}}

\caption{Temperature dependence of the O K-edge spectra for the a) 10 and b) 20~nm Co$_3$O$_4$ nanocubes.\label{fig:O_XAS_vs_T}  c) Temperature dependence of $t_{\text{2g}}/e_{\text{g}}$.  The inset shows the temperature dependence of the O 2p ligand hole indicated by the pre-peak intensity as identified in Fig.~\ref{fig:O_XAS}.}

\end{figure*}


The EXAFS oscillations in the $\beta$ and beyond energy regime ($\sim$7730 to 7750~eV in figure~\ref{fig:XANES}) were quantified by first fitting a cubic spline function through the post-edge region of the XAFS using ATHENA\cite{XAFS_Artemis}, and then modelled and fitted using ARTEMIS\cite{XAFS_Artemis}.  To help identify the salient features, we present $k$-space wavelet transforms of the EXAFS spectra for the 10~nm and 20~nm Co$_3$O$_4$ nanocubes calculated using a Morlet wavelet with a {$k$}-weighting of 2 (figure~\mbox{\ref{fig:EXAFS_10nm}} and figure~\mbox{\ref{fig:EXAFS_20nm}}). The resulting wavelet transforms were compared with that of a reference bulk Co$_3$O$_4$\cite{xasdb_nist} (see Supporting Information, figure~S10). The wavelet parameters were fixed at \mbox{$\sigma = 1$} and \mbox{$\kappa = 10$} for all three datasets. Overall, the wavelet transforms of both nanocube samples closely resemble that of the reference Co$_3$O$_4$, indicating that the local coordination environment is largely preserved. The feature centred at approximately 1.45~\AA~corresponds to the Co-O single-scattering path, while the higher-$R$ feature between 2.5 and 3.0~\AA~arises from Co-Co scattering paths. The wavelet transform of the 10~nm Co$_3$O$_4$ nanocubes is very similar to that of the reference sample, with the exception of a reduced intensity near $R \approx 3$~\AA~ and $k \approx 10$~\AA$^{-1}$. This reduction may correspond to a lower Co-Co coordination number. In contrast, the Co-O feature of the 20~nm Co$_3$O$_4$ nanocubes exhibits noticeably greater distortion than that of both the reference and the 10~nm nanocubes, suggesting a reduction in the Co-O coordination. Furthermore, the wavelet transform features of both nanocube samples are broader than those of the reference, indicating increased structural disorder. This broadening is more pronounced for the 20~nm nanocubes, implying a higher degree of local disorder than in the 10~nm nanocubes.

The real-space representation of the EXAFS spectra are shown on the right side in figure~\mbox{\ref{fig:EXAFS_10nm}} and figure~\mbox{\ref{fig:EXAFS_20nm}}, where the solid lines through the data are fits confined between $k=3$ and 13~\AA$^{-1}$~using a Hanning window. The data were fitted in real space from 1 to 3.6~\AA~using $k$-weights of 1, 2, and 3. Theoretical calculations of phase shifts and backscattering amplitudes for the coordination atomic pairs were obtained using FEFF \mbox{\cite{FEFF}} with the crystallographic data taken from the refinements of the XRD patterns (discussed above). The fits are shown by the solid lines in Figures~\ref{fig:EXAFS_10nm} and \ref{fig:EXAFS_20nm}.  Supporting Information Table~S1 provides the fit results and they are entirely consistent with Co$_3$O$_4$\cite{Mijiti.2021, Bai.2012, Bergmann.2015, Saddeler.2021, Gawai.2022, Wang.2016}.  We find that the nanocubes have essentially identical bond distances between the Co$^{2+/3+}$ (O$_{\text{h}}$), Co$^{2+}$(T$_{\text{d}}$) and O atoms in their spinel structures (i.e. same bond distances, $R$). The only significant difference is that the 20~nm nanocubes are much more structurally disordered, exemplified by their larger and less certain Debye-Waller factors ($\sigma^2$) and less well defined scattering path multiplicities ($N$).

\subsection{Surprising valence and coordination changes}

This structural disorder has an impact on the Co coordination environments, as revealed by the Co L$_{\text{3,2}}$-edge x-ray absorption spectroscopy (XAS) shown in figure~\ref{fig:Co_XAS}.  Ligand-field simulations (parameters in Supporting Information Table~S2\cite{Shepit.2020, Wang.2022, Kumar.2022}) of the low-spin O$_{\text{h}}$ Co$^{3+}$ and T$_{\text{d}}$ Co$^{2+}$ identify that essentially half of each coordination is present in the 10~nm nanocubes (solid line in figure~\ref{fig:Co_XAS_10nm}), quite different from the 2/3 O$_{\text{h}}$ Co$^{3+}$-to-1/3 T$_{\text{d}}$ Co$^{2+}$ of bulk Co$_3$O$_4$.  The XAS data for this system also exhibits a `pre-peak' just before the L$_{\text{3}}$-edge at $\sim 777$~eV that is not all at described with the usual cations.  O$_{\text{h}}$ Co$^{2+}$ (pink solid line in figure~\ref{fig:Co_XAS_10nm}) does correlate with that peak position, and a site-weighted fit with all three coordinations results in the description shown by the solid pink line through the data for ~$21\pm1$\% Co$^{2+}$ O$_{\text{h}}$, $35\pm1$\% Co$^{3+}$ O$_{\text{h}}$ and $44\pm1$\% Co$^{2+}$ T$_{\text{d}}$.  By contrast, the 20~nm Co$_3$O$_4$ nanocubes XAS is well described by the more conventional Co$^{3+}$ O$_{\text{h}}$ and Co$^{2+}$ T$_{\text{d}}$ coordinations at $58\pm1$\%-to-$42\pm1$\%, respectively.  Interestingly, including the odd Co$^{2+}$ O$_{\text{h}}$ coordination in the 10~nm nanocubes, the total O$_{\text{h}}$-to-T$_{\text{d}}$ ratios are the same between the two sizes.  This overlap and its $\sim10$\% change from the bulk points to overall disorder (structural and chemical via perhaps Co vacancies) from finite-size effects, e.g. these results could be indicative of $\sim30$\% O$_{\text{h}}$ Co sites vacancies -- indeed, the weak positive slope of the XAS intensity for energies \textgreater 800~eV instead of the usual return to zero (e.g. for a metal) is a sign of shared Co d-electrons with the O s-p electrons introduced by Co vacancies, and in our experience typical of nanoscale transition metal oxides.

To help identify what may be affecting the Co ions and if vacancies are present (Co and O), we turn to O K-edge XAS, shown in figure~\ref{fig:O_XAS}.  These spectra are a measure of the transitions for O 1s core levels into empty O 2p states that are hybridized with the Co d-orbital electrons, thus the spectral peaks are associated with these empty states.  Peak intensities and energy shifts quantify 2p states mixed with the Co and if extra (or missing) electrons are due to altered oxidation states.  Any extra electrons will be trapped near vacancies and create (or alter) spectral features for these new, localized states below the conduction band.  Specifically, the region between 529 and 533~eV shows two peaks due to transitions from the O 1s state to hybridized O 2p–Co 3d unoccupied states, and the region at higher energies contains a broad multi-peak structure that is due to transitions from the O 1s states to O 2p–Co 4s and 4p hybridized states\cite{Frati.2020, Kenmoe.2022}.  The presence of new states is made clear by the two peaks between 530 to 533~eV that are a direct measure of the t$_{\text{2g}}$ and e$_{\text{g}}$ electron populations about the O ions in the Co$_3$O$_4$ nanocubes.  In comparison, for bulk Co$_3$O$_4$ these peaks are unresolved, and a single peak at 531~eV is present\cite{Kenmoe.2022, Wu.2026}.  The relative intensities of the spectral peaks at $\sim 530$~eV (t$_{\text{2g}}$ states) and $\sim 531$~eV (e$_{\text{g}}$ states) reflect that the 20~nm Co$_3$O$_4$ nanocubes Co (O$_{\text{h}}$) are more hybridized with the O 2p bands than the 10~nm nanocubes (more on this below).  Furthermore, the highly delocalized electronic environments due to hybridization between the O and Co in the nanocubes is qualified by the washed out spectral features between 535 and 550~eV when compared to bulk Co$_3$O$_4$\cite{Kenmoe.2022, Wu.2026}.  Most telling is a comparison of the spectral features associated with the population of localized O 1 s electrons (their densities) to those electrons hybridized with the Co ions (i.e. the relative intensities of the region in orange in figure~\ref{fig:O_XAS} to the 535 to 545~eV region \cite{misc_comm2} \nocite{Lin2014_surface}).  In bulk Co$_3$O$_4$ these energy regions have comparable intensities \cite{Kenmoe.2022, Wu.2026}. The Co$_3$O$_4$ nanocubes exhibit the opposite behaviour (indication significant delocalization and hybridization\cite{Frati.2020}) that is dramatically size dependent (figure~\ref{fig:O_XAS} orange region) and a measure of vacancies about the Co ions.  Additionally, there is a new (compared to the bulk) spectral feature between 527 and 529~eV (inset in figure~\ref{fig:O_XAS}) that is also size dependent that further demonstrates that presence of O vacancies.  This spectral feature indicates the presence of 2p ligand core holes that form due to a shift in the electron density from the O 2p towards the Co (O$_{\text{h}}$) d states. There seems to be a redistribution of charge along the Co$^{3+}$ (O$_{\text{h}}$)-O bonds, consistent with a negative charge transfer energy (Table~S2 and S6) that make the $3d^7\underline{L}$ ground state more stable than the ionic 3d$^{6}$ state for the Co$^{3+}$ (O$_{\text{h}}$) sites.  This stronger hybridization leads to stronger charge transfer values\cite{Wang.2022} that we observe in our Co $L_{3}$-edge ligand-field simulations, discussed below.

10~K Co L$_{\text{3,2}}$ edge XMCD of the 10 and 20~nm Co$_3$O$_4$ nanocubes are shown in figures~\ref{fig:Co_XMCD_10nm_10K} and~\ref{fig:Co_XMCD_20nm_10K}, respectively.  These spectra were collected using right and left circular polarizations (\textgreater 95\%) and $\pm4$~T to remove any artifacts from polarization differences and charge effects (e.g.~TEY detector and charge buildup during the scans).  Total flourescence yield (TFY) XMCD were also collected (not shown) and present the same spectral lineshapes.  The effects to the nanocubes Co and O valences and coordiantions from finite size effects and disorder are definitely revealed; there is an XMCD signal; part of the Co in the nanocubes is ferromagnetically coupled.  Recall that Co$_3$O$_4$ is usually a fully spin compensated antiferromagnet, so there are equal Co spin up and spin down populations and thus no elemental magnetization to measure.  In its bulk form, Co$_3$O$_4$ has Co$^{2+}$ (T$_{\text{d}}$) and Co$^{3+}$ (O$_{\text{h}}$) in a 1:2 ratio (Supporting Information Table~S3), with Co$^{3+}$ (O$_{\text{h}}$) are non-magnetic as the electrons pair-up in the lower t$_{\text{2g}}$ orbitals due to a large crystal field splitting, so that only the Co$^{2+}$ (T$_{\text{d}}$) are are magnetic due to their comparatively lower crystal field splitting.  These Co$^{2+}$ (T$_{\text{d}}$) are antiferromagnetically coupled and fully compensated in the spinel structure\cite{Roth.1964}.

Clearly, the above electronic structure (that results in the unusual nanomagnetism described above) has been altered in the Co$_3$O$_4$ nanocubes.  Ligand-field multiplet simulations only succeed in capturing the overall XMCD lineshape (solid orange lines in figures~\ref{fig:Co_XMCD_10nm_10K} and~\ref{fig:Co_XMCD_20nm_10K}) using the complete range of possible Co valences and coordinations, with simulation parameters provided in Supporting Information Table~S2. The corresponding XAS spectra were simulated again after taking into account all the Co ions identified from the XMCD analysis. The resulting relative occupancies of Co ions are listed in the Supporting Information Table~S3. From these results, we find that the ratio of T$_{\text{d}}$ to O$_{\text{h}}$ Co is less than in the bulk (52\%) compared to 19\% and 16\%for the 10 and 20~nm nanocubes, respectively, that indicates a large number of T$_{\text{d}}$ sites vacancies. The presence of these vacancies can promote O vacancies that introduce additional Co$^{2+}$ (O$_{\text{h}})$\cite{Frati.2020}. This observation is consistent with the larger lattice constants measured (discussed above).  Also, the O vacancies further decrease the crystal fields around the Co$^{3+}$ so that they stabilize in the HS state\cite{Long.2025}, as we observe.  This significant reduction in Co (T$_{\text{d}}$) coordination is also echoed in the O K-edge XAS spectra discussed above.

Any changes in stress or strain with temperature that could be responsible for the unusual nanomagnetism identifed by the 10~K XMCD spectra should reveal themselves in lineshape changes in the XAS, and Supporting Information figure S4 shows that the overall Co$^{2+}$ and Co$^{3+}$ valence and coordination in both nanocube systems remains unchanged from 10 to 200~K.  By contrast, the temperature dependent XMCD of the Co L$_{\text{3}}$-edge (Fig.~Supporting Information S5) presents a definite evolution of the magnetic spin configurations with increasing temperature and it is different between the 10 and 20~nm Co$_3$O$_4$ nanocubes. Despite comprising only about 17-20 \% of the total Co obtained from XAS, the observed magnetism comes mostly from the Co$^{2+}$(O$_{\text{h}}$) (dashed line in figure~S5). This is because the other Co ions are participating in antiferromagnetic exchange and therefore do not contribute to XMCD signal, whereas Co$^{2+}$(O$_{\text{h}}$) ions are (more likely to be) ferromagnetically coupled or compensated. Moreover, Co$^{2+}$(O$_{\text{h}}$) ions have a larger contribution from an unquenched orbital moment. Figure S5 shows an evolution of the high-spin and low-spin configurations with warming. The changes in the relative occupancies of different sites from Supporting Information Tables~S4 and S5 reveal a progression of hybridization and spin exchange between Co and O ions with increasing temperature. The temperature dependence of the Co$^{2+}$(O$_{\text{h}}$) magnetization from the XMCD, shown in figure~\ref{fig:M_Co_XMCD}, parallels $M_{\text{AFM}}(T)$ below $\sim50$~K (figure~\ref{fig:MAFM_vs_T}), a strong indication that the elemental magnetization measured by the XMCD is from the nanocube surface spins.

As most of the magnetism observed in the Co $L_{3}$-edge XMCD is from the Co(O$_{\text{h}}$) ions using the assignments provided by the ligand-field simulations, we find that the Co 3d levels have been split into lower energy $t_{\text{2g}}$ orbitals, while the higher-energy 3d levels move towards $e_{\text{g}}$ orbitals. Consequently, in the O K-edge spectrum shown in figure~\ref{fig:O_XAS_vs_T}, the lower-energy pre-peak represents the $t_{\text{2g}}$ states while the higher-energy pre-peak represents the $e_{\text{g}}$ states of electrons of the O ions whose covalency is affected. \cite{misc_comm1} \nocite{Wang2022Co3O4RIXS}  Indeed, the temperature evolution of the $t_{\text{2g}}$ and $e_{\text{g}}$ spectral features occur in lock-step with the spectral changes exhibited by the Co $L_{\text{3}}$ XMCD (figure~S5 and Tables~S4 and S5), further strengthening the link between the electronic changes with temperature occurring between Co and O throughout the Co$_3$O$_4$ nanocubes and the resultant nanomagnetism. This is attributed to the temperature induced distortions revealed by the EXAFS of the Co-O bonds that influence the O (2p) - Co (3d) hybridization differently for the $t_{\text{2g}}$ and $e_{\text{g}}$ orbitals (figures S6, S7, S8 and S9). As the intensity of these spectral peaks is a measure of the `d-character' mixed with the O 2p orbitals, where $t_{\text{2g}}$ orbitals are directed between ligands and $e_{\text{g}}$ orbitals are directed at the ligands.  The $t_{\text{2g}}$-to-$e_{\text{g}}$ ratio temperature dependence (figure~\ref{fig:t2geg_vs_T}) shows how the Co-O ligands stabilize and localize more towards the usual Co$_3$O$_4$ Co(O$_{\text{h}}$) low-spin coordination with increasing temperature.  This trend is also observed in by the vanishing O 2p ligand holes as electrons move back to reestablish the Co$^{3+}$(O$_{\text{h}}$) low-spin states.


\begin{figure}[t!]

\centering

\includegraphics[width=0.41\textwidth]{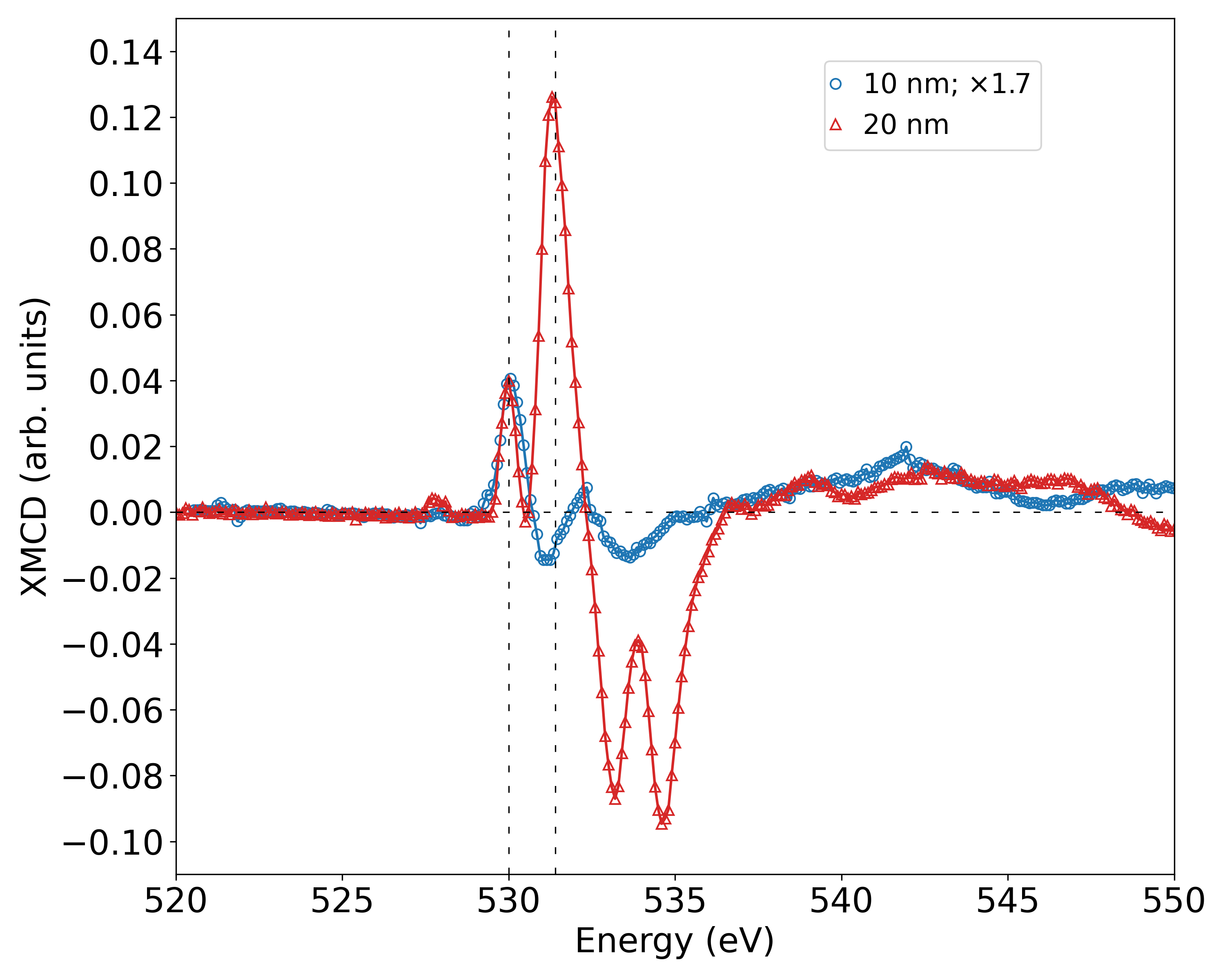}

\caption{Oxygen K-edge XMCD at 10~K for the 10~nm ($\bigcirc$; blue) and 20~nm ($\bigtriangleup$; red) Co$_3$O$_4$ nanocubes.  Note that the 10~nm XMCD signal has been multiplied by a factor of 1.7. The dashed lines mark the $t_{\text{2g}}$ and $e_{\text{g}}$ transitions from the XAS.\label{fig:O_XMCD}}

\end{figure}


\subsection{Unanticipated nanomagnetism with frustration}

Most fascinating is the observation of magnetic O in the Co$_3$O$_4$ nanocubes, as revealed by the 10~K XMCD on the O K-edge shown in figure~\ref{fig:O_XMCD}.  The $t_{\text{2g}}$ and $e_{\text{g}}$ electrons have hybridized to the extent that there are clearly (overall) unpaired spin populations and they are antiferromagnetically coupled to the Co$^{2+}$ and Co$^{3+}$ spins in all configurations, with the counterinfluence of the Co$^{2+}$(O$_{\text{h}}$) low-spin versus high spin states identified by the opposite alignment of the $e_{\text{g}}$ spins in the 10~nm nanocubes compared to the 20~nm nanocubes respectively. Note that Co$^{2+}$(O$_{\text{h}}$) in the low spin state has additional unoccupied majority spin channels in addition to the miniority spin channels in the unoccupied $e_{\text{g}}$ orbitals, compared to the presence of only unoccupied miniority spin channels in the unoccupied $e_{\text{g}}$ orbitals of Co$^{2+}$(O$_{\text{h}}$) in the high spin state.  The spectral features observed around 534-535~eV likely originate from the hybridization of O (1s) orbitals with Co (3d) orbitals that are shifted to higher energies as a result of distortions in the Co-O bonds of the nanocubes. Evidence for such bond distortions is shown by the wavelet transform analysis (figures~\ref{fig:EXAFS_10nm}, ~\ref{fig:EXAFS_20nm}, S7, and S9) that show distortions of the Co-O scattering features toward higher $k$-values. These distortions are more pronounced in the 20~nm nanocubes, consistent with the substantially larger XMCD signal provided by the 20~nm sample.  The magnetic O ions identified by this XMCD are predominately from the nanocubes' surfaces, as is revealed by the scaling of the O magnetization of the 10~nm with respect to the 20~nm nanocubes (i.e. (17~nm)$^2$/(13~nm)$^2 = 1.7$).  This O XMCD weakens considerably with increasing temperature and is not observable by $\sim50$~K (not shown).


\begin{figure*}

\centering

\subfloat[a) \label{fig:spin_calculations}]{\includegraphics[width=0.375\textwidth]{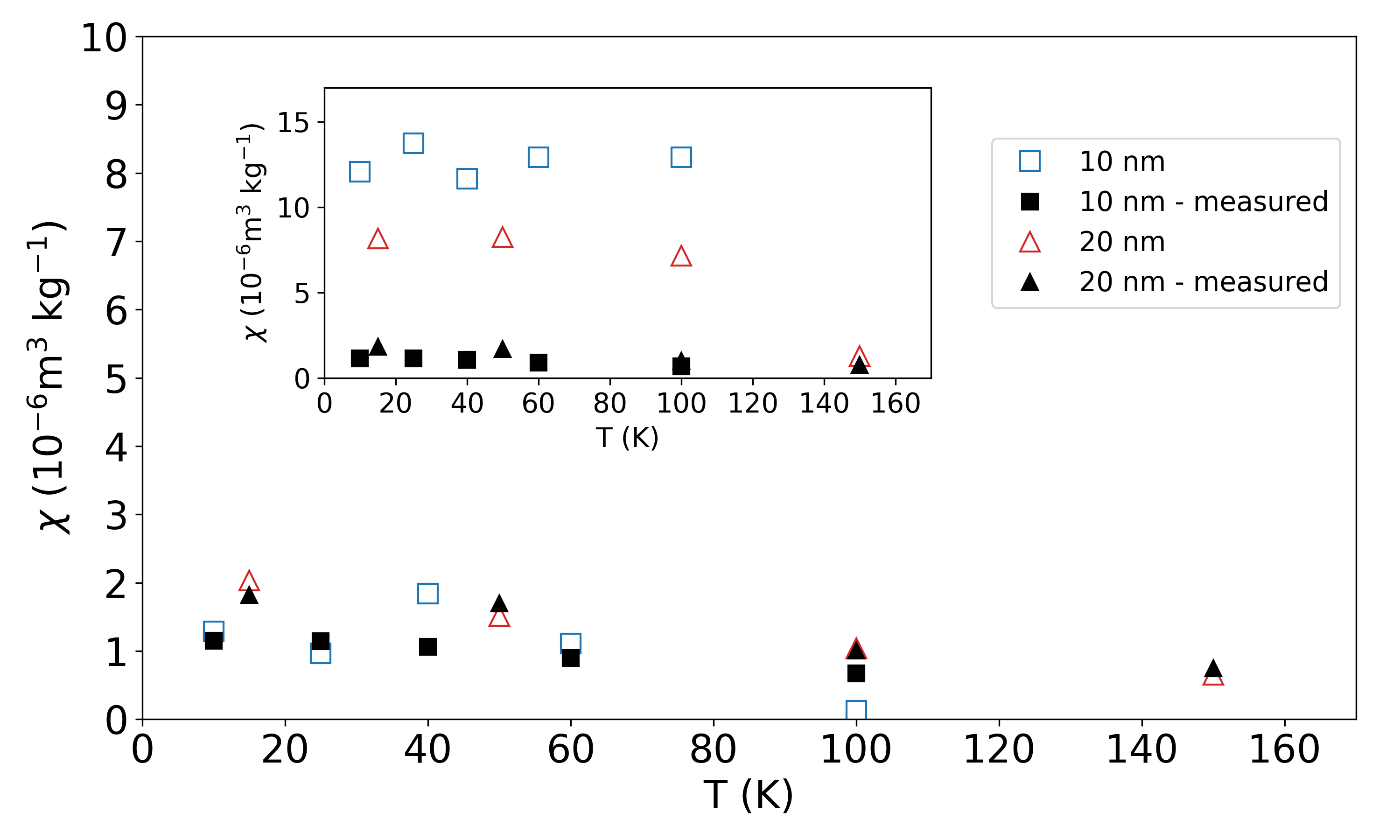}}
\hfill
\subfloat[b) \label{fig:Xac_vs_T_10nm}]{\includegraphics[width=0.275\textwidth]{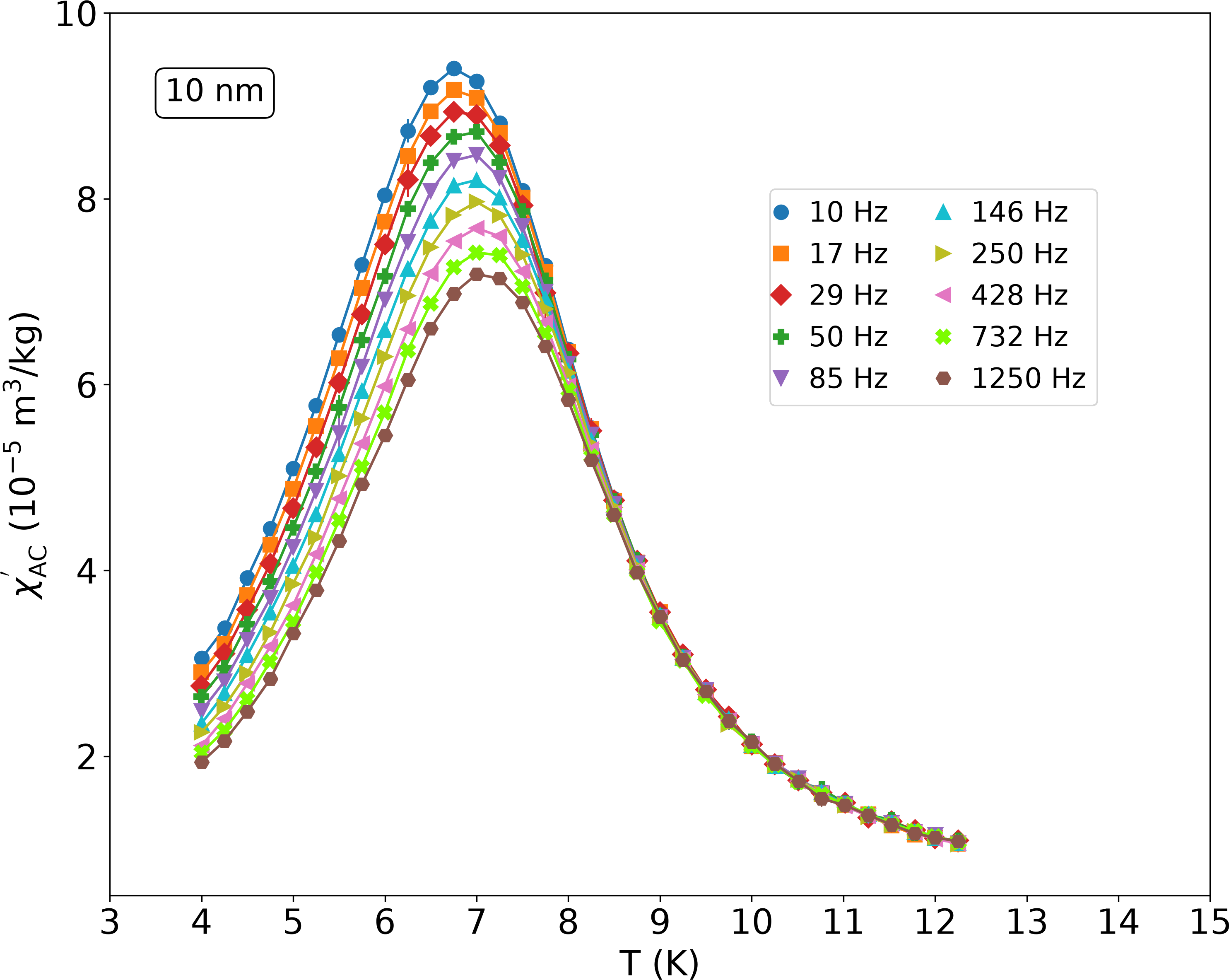}}
\hfill
\subfloat[c) \label{fig:VF_fit}]{\includegraphics[width=0.275\textwidth]{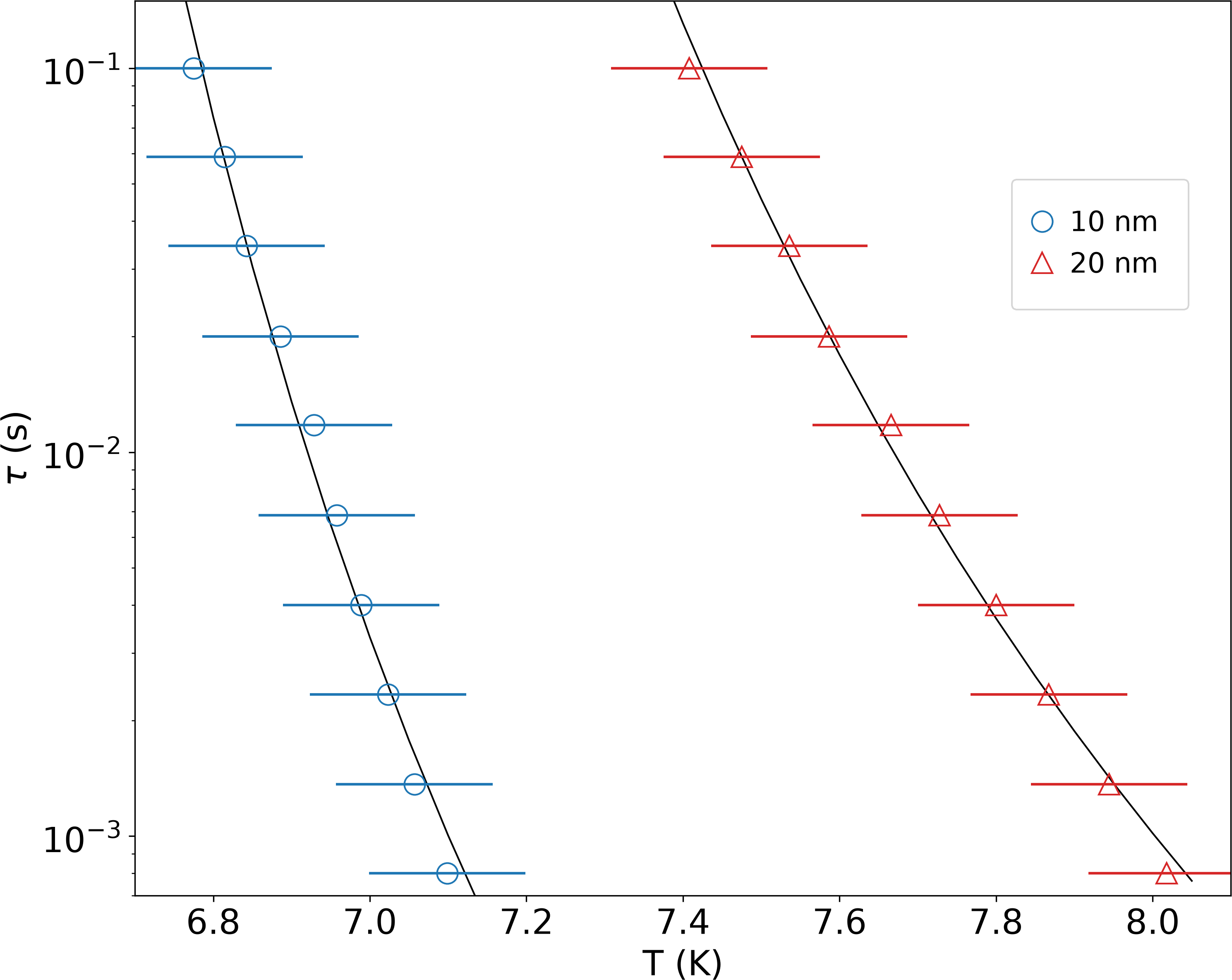}}

\caption{a) A comparison between the calculated $\chi$ and the experimental linear high field $\chi_{\text{DC}}(T)$ of the 10 nm ($\Box$) and 20 nm ($\bigtriangleup$) Co$_3$O$_4$ cubes. The open symbols represent the calculated values and the closed black symbols represent the respective experimental values. The inset shows comparison of the calculated $\chi$ without O$^{2-}$ ions and the experimental values.  b) Temperature and frequency dependence of the real part of the AC susceptibility, $\chi_{\text{AC}}(\nu,T)$ for the 10~nm nanocubes.  c) Measuring time ($\tau=1/\nu)$ dependence of the cusp (freezing) temperature from $\chi_{\text{AC}}(T)$.  Solid lines are fits described in the text.\label{fig:Xac_vs_T}}

\end{figure*}


To relate all the electronic and spin configuration information with the overall measured nanomagnetism, we map the measured $\chi_{\text{HF}}(T)$ (figure~S3b) to a spin calculation considering all possible moment sources (as described in the Supporting Information. A theoretical susceptibility is calculated by considering the core spinel structure and the contributions from the changed stoichiometries on the surfaces of Co$_3$O$_4$ cubes.  Starting with the consideration that the entire magnetic response is from the surface (the core is antiferromagnetic) and thus from Co ions in the above described oxidation and spin states (from O$_{\text{h}}$ and T$_{\text{d}}$ ligands), we find there is a large discrepancy between experiment and theory; a large overestimate, shown in figure~\ref{fig:spin_calculations}). This overestimate is accounted for by including O$_2^{-}$-like species due to the broken bonds and vacancies\cite{Preda.2011}. The presence of O$_2^{-}$-like species is related directly to the presence of polarized Co$^{3+}$-O 2p hybridized bonds, also evident from O K-edge XAS analysis. Therefore, from the discussion so far, the presence of magnetic frustration can  be attributed to competition between the polarized Co-O bonds, and highly anisotropic Co$^{2+}$, and Co$^{3+}$ high-spin ions results in the observed magnetic frustration.

The remaining question is what kind of nanomagnetism is responsible for $M_{\text{AFM}}(T)$: the surface spins.  No measurable coercivity would indicate no coherent (cooperative) alignment of spins and no surface anisotropy, a property characteristic of magnetic nanoparticles\cite{Coey.2019,Liu.2025, Wesselinowa.2010}.  But, surface spins typically experience magnetic frustration, and the necessary ingredients for magnetic frustration are  listed and quantified by the measured structural and chemical disorder as well as the complex interrelationship between spin configurations and temperature elucidated by the XAS and XMCD results.  The hall-mark signature of magnetic frustration is a spin-glass-like cusp in $\chi_{\text{DC}}(T)$, but while the measurements hint at the emergence of a cusp with cooling (figure~S11), more conclusive results are needed.  Realizing that the spin-glass cusp is due to dynamical freezing of spins that can be quantified by a measuring-time dependence of the magnetic response (low-field susceptibility to avoid complications inherent with larger applied fields), we turn to frequency and temperature dependent AC susceptibility measurement, shown in figure~\ref{fig:Xac_vs_T}.  Using faster measuring times (compared to $\chi_{\text{DC}}(T)$) reveals the signature cusp of the susceptibility (spin response) with clear measuring time ($\tau$) (or frequency ($\nu$)) dependencies for both 10 and 20~nm nanocubes.  The cusp temperature shifts to higher temperatures with shorter measuring times which is why the cusp appears with measuring times faster than \textgreater 1~s (the time scale of the $\chi_{\text{DC}}(T)$ measurements).  These measured (monotomic) amplitude and measuring time dependencies of $\chi_{\text{AC}}(\nu,T)$ are the hallmarks of dynamical freezing from frustrated spins in a spin-glass-like configuration.  Indeed, using the Mydosh parameter\cite{Mydosh.book}, $\delta_f = (\Delta T_f)/(T_f \Delta\log_{10}\{{\nu}\})$, where $T_f$ is the cusp (spin freezing) temperature, we find $\delta_f = 2.2\pm0.01\times10^{-1}$ and $3.9\pm0.01\times10^{-1}$ for the 10 and 20~nm Co$_3$O$_4$ nanocubes, respectively.  These results are in good agreement with cluster-spin-glass\cite{Smith.1974,Mydosh.2015} behaviour and well outside the range of $\delta_f$ values associated with single-domain nanoparticle superparamagnets.  A cluster-spin-glass has the disorder and frustration among regions of a small number of exchange-coupled spins, a representation fully consistent with all the results discussed above.

Further evidence of the measured glass-dynamics is provided by the time (frequency) dependence of $T_f$ (figure~\ref{fig:VF_fit}) that is well described by Vogel-Fulcher-Tammann\cite{Mydosh.book, Mydosh.2015, Tholence.1980} $\tau = \tau_0 \exp{(E_a/\{k_B(T_f - T_0)\})}$, where $\tau_0$ is a spin-reversal attempt time over an energy barrier $E_a$ that is affected by an `interaction energy' $T_0$.  Fits (solid lines in figure~\ref{fig:VF_fit}) yield $\tau_0 = 10^{-9 \pm 1}$~s, and $E_a = 24\pm1\times10^{-24}$~J and $37\pm6\times10^{-23}$J for the 10~nm and 20~nm Co$_3$O$_4$ nanocubes, respectively, values that are reasonable for a cluster-glass-system.  $T_0 is 5.8\pm0.2$~K for the 10~nm nanocubes, and $5.84\pm0.05$~K for the 20~nm nanocubes.  The essentially identical $E_a$ and $T_0$ values indicate that the spin-cluster sizes (regions where spins are exchange coupled) are unaffected by nanocube size, and that underlying mechanisms leading to the freezing processes are different and controllable via the oxygen-driven nanomagnetism described above.  Specifically, the differences in hybridized bonds and their covalencies that permit the O magnetization (figure~\ref{fig:O_XMCD}) between the 10 and 20~nm Co$_3$O$_4$ nanocubes pushes the spin freezing temperature higher and offers a unconventional way to turn an antiferromagnetic nanomaterial into an artificial spin-glass nanosystem, providing application benefits of antiferromagnetism and the hierarchical energy landscape of a spin-glass.

\section{Summary and perspectives}

Co$_3$O$_4$ nanocubes that expose (100) surface terminations have modified coordination between respective Co ions such that the polarized Co-O bonds compete with anisotropic free spins (Co$^{2+}$ O$_{\text{h}}$ and high-spin Co$^{3+}$ O$_{\text{h}}$) resulting in an artificial spin glass nanosystem. Unlike conventional antiferromagnet nanomagnetic systems where a modified magnetic structure is confined to the surface (arising from exchange-couple-free surface spins interacting with a partially or fully spin-compensated core), our nanocube systems reveal the presence of dramatically altered electronic structures around ions throughout the particle. This is clear from high field and low field magnetic DC susceptibility and $M$ vs $\mu_0H$ of the nanocubes compared with 400~nm cube structures. The magnetic measurements link the modified exchange pathways with the presence of oxygen vacancies and Co$^{2+}$ $O_{\text{h}}$ leading to spin freezing at low temperatures. EXAFS studies identify the distorted Co-O bonds and reduced Co-Co coordinations that are more pronounced in the 20~nm Co$_3$O$_4$ cubes than the 10~nm Co$_3$O$_4$ cubes.  Further, Co L$_{\text{3,2}}$-edge XAS and XMCD studies revealed the `missing' Co ions responsible for O vacancies, and additional Co$^{2+}$ and high-spin Co$^{3+}$ ions in O$_{\text{h}}$ sites. All together, these results identify a very different (from conventional antiferromagnetic) Co$_3$O$_4$ electronic structure that has high spin Co$^{2+}$ in T$_{\text{d}}$ sites and low spin Co$^{3+}$ in O$_{\text{h}}$ sites. The temperature dependent XAS and XMCD also quantified the contributions of the new ions in O$_{\text{h}}$ sites to the unusual nanomagnetism and its evolution with temperature. The temperature dependent O K-edge XAS showed the changes in the Co-O hybridization and revealed unexpected polarized Co-O bonds due to O 2p ligand core holes which are dominant in the low temperature region. Additionally, the changes in Co-O hybridization has resulted in unusual O K-edge XMCD which is largely from the surface as revealed from the difference in the XMCD signal between 10~nm and 20~nm nanocubes. Finally, AC susceptibility measurement and analysis revealed the formation of an artificial spin-glass like magnetic order that is considered to arise from polarized Co-O bonds and free  Co$^{2+}$ O$_{\text{h}}$ and high spin  Co$^{3+}$ O$_{\text{h}}$ ions.

For applications, conventional spintronics magnetic nanomaterials have the a magnetization from ferromagnets that make detection and manipulation straightforward.  Contrariwise, antiferromagnetic nanomaterials are potentially much faster since they do not present a net magnetization.  Herein, we show how antiferromagnetic nanosystems can be built to perform spin-glass magnetism that can be tuned in unforeseen ways via i) size and shape, and ii) through unique surface environments and their impact on \emph{both} oxygen and transition metal ion electronic environments.  Further improvements (e.g. $T_f$  above room temperature) should be possible by way of astute doping, careful size control, and designing heterostructures like Fe$_3$O$_4$/Co$_3$O$_4$ core/shell nanocube systems (that show similar enhancements to their nanomagnetism as their nanosphere counterparts, e.g. Ref.\cite{Skoropata.2014, Oberdick.2018}). Indeed, our preliminary investigations following these avenues are yielding promising results.  Perhaps most exciting is that our artificial spin-glass of antiferromagnetic nanocubes is a potentially interesting platform for physical reservoir computing\cite{Vahl.2024} and active inference devices\cite{Stamps.2025}.  Intrinsically, spin-glasses are a natural framework for physical reservoir computing.  What makes spin-glasses challenging to explore and understand -- metastability, a wide range of spin relaxation rates, non-linearity, disorder, frustration, and (most importantly) a plethora of interacting states -- are the same properties that make reservoir computing revolutionary\cite{Vahl.2024, Heins.2023, Saccone.2022, Tanaka.2019}.  The required nonlinear responses to inputs and a fading memory, and a high dimensional of internal states, are all parts of spin-glass behaviour.  Herein, we have demonstrated that our artificial spin glass presents the requisite nonlinearity and high-dimensional dynamics that can be field-controlled with a measurable output.  These behaviours can be enhanced via tuned (nonlinear) spin precession, interparticle interactions and thermal fluctuations.  Preliminary investigations reveal the necessary fading memory of this artificial spin glass and the multitude of reservoir states, and we are currently validating the capability of our artificial spin glasses for physical reservoir computing.  The antiferromagnetic Co$_3$O$_4$ nanocubes should also present very high frequency spin dynamics that would allow computing speeds difficult to achieve with conventional (ferro)magnetic systems -- a potential pathway to high-speed physical computing\cite{Tanaka.2019, Chen.2026}.



\bmsubsection*{Acknowledgments}
The authors (G.K.U. and J.v.L.) acknowledge funding from the Natural Sciences and Engineering Research Council of Canada (RGPIN 2024-04882 and ALLRP 580243-22) and the Canada Foundation for Innovation.  Use of the Advanced Photon Source at Argonne National Laboratories, an Office of Science User Facility operated for the U.S. Department of Energy (DOE) was supported by the US DOE under contracts DE-AC02-06CH11357 and DE-AC02-05CH11231. Use of the Advanced Light Source at Laurence Berkeley National Laboratory, a U.S. Department of Energy (DOE) Office of Science User Facility was supported by the U.S. DOE under Contract No. DE-AC02-05CH11231.  The authors (R.N, G.K.U., J.v.L.) thank the European Synchrotron Radiation Facility for providing beamtime under proposal IHHC4164.

\bmsubsection*{Conflicts of Interest}

The authors declare no conflicts of interest.

\bmsubsection*{Data Availability Statement}

The data is available upon reasonable request.    The raw absorption data are available at \url{doi.org/10.15151/ESRF-ES-2157153546}; embargo periods may apply.

\bibliography{Co3O4_nanocubes_AFM_GKU.bib}

\begin{thebibliography}{10}
\bibitem{misc_comm1}
 An alternative interpretation of the two pre-edge features could be in terms
  of majority- and minority-spin states. However, this assignment is less
  consistent with the expected electronic structure of Co$_3$O$_4$ because for
  the Co oxidation states and spin configurations considered to describe XAS
  and XMCD, the unoccupied Co 3d states that hybridize with the O 2p states are
  predominantly associated with the minority-spin channels. Thus, the observed
  separation of the two pre-edge features is more naturally explained by the
  crystal-field-derived electronic structure of the inequivalent Co sites than
  by a simple majority- and minority-spin splitting.
\end{thebibliography}

\bmsubsection*{Supporting Information}

Additional supporting information can be found online in the Supporting Information
section.

\bmsubsection*{Table of Contents}
\begin{figure}[h!]

\centering

\includegraphics[width=0.30\textwidth]{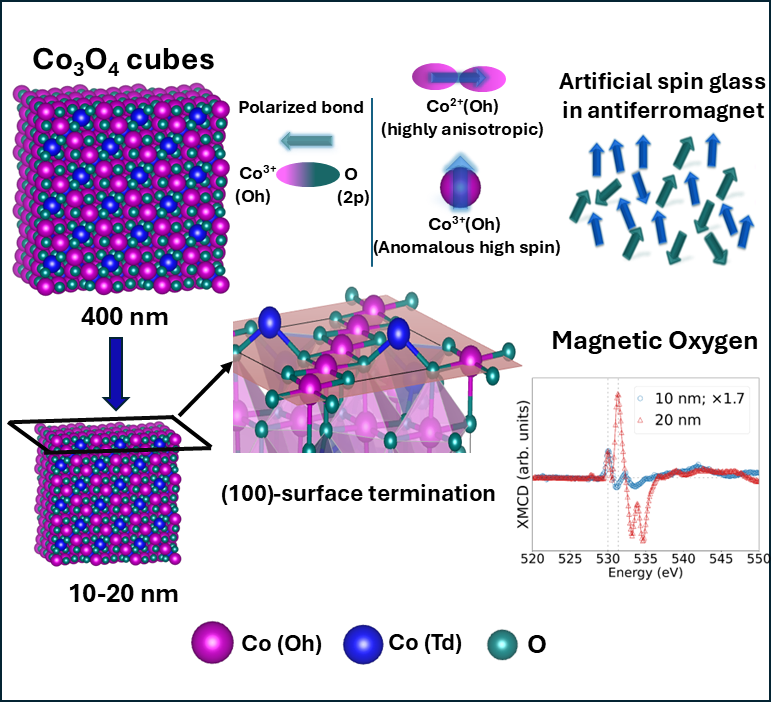}

\caption{}

\end{figure}
Co$_3$O$_4$ nanocubes exposing (100) surfaces exhibit an altered ionic arrangement compared with the bulk antiferromagnetic structure, producing an unusual spin-glass-like magnetism and magnetic oxygen. This magnetic frustration can be tuned by controlling nanocube size, offering a new platform for studying nanomagnetism and advancing technologies such as spintronics and physical reservoir computing.


\end{document}